\documentclass[a4paper,11pt]{article}
\usepackage{jcappub} % for details on the use of the package, please
\usepackage[T1]{fontenc} % if needed
\RequirePackage{amsfonts,bm,hhline,amsmath,amssymb,float,mathtools,microtype,eurosym,latexsym,epsf,times,makecell,array,natbib}
\usepackage{hyperref}
\usepackage{epstopdf,color}
\usepackage[font={footnotesize,it}]{caption}
\usepackage[caption = false]{subfig}
\setcellgapes{4pt}%parameter for the spacing

\newcommand*\DAlambert{\mathop{}\!\mathbin\Box}
\title{\boldmath Study on charged strange stars in $f\left(R,\mathcal{T}\right)$ gravity}

\author[a]{Debabrata Deb,}
\author[b]{Sergei V. Ketov,}
\author[c]{Maxim Khlopov}
\author[d,1]{Saibal Ray\note{The corresponding author.},}

\affiliation[a]{Department of Physics, Indian Institute of Engineering Science and Technology, Shibpur, Howrah 711103, West Bengal, India}
\affiliation[b]{Department of Physics, Tokyo Metropolitan University, 1-1 Minami-ohsawa, Hachioji-shi, Tokyo 192-0397, Japan \\ \&  Research School of High-Energy Physics, Tomsk Polytechnic University, 2a Lenin Avenue, Tomsk 634050, Russian Federation \\ \& Kavli Institute for the Physics and Mathematics of the Universe (WPI), University of Tokyo, Kashiwa 277-8583, Japan}
\affiliation[c]{APC Laboratory 10, rue Alice Domon et L{\'e}onie Duquet, 75205 Paris Cedex 13, France  \\ \&  Institute of Physics, Southern Federal University, 194 Stachki, Rostov-on-Don 344090, Russian Federation}
\affiliation[d]{Department of Physics, Government College of Engineering and Ceramic Technology, Kolkata 700010, West Bengal, India}

\emailAdd{ddeb.rs2016@physics.iiests.co.in}
\emailAdd{ketov@tmu.ac.jp}
\emailAdd{khlopov@apc.in2p3.fr}
\emailAdd{saibal@associates.iucaa.in}

\abstract{We investigate effects of the modified $f(R, \mathcal{T})$ gravity on the charged strange quark stars with the standard choice of $f(R, \mathcal{T})=R+2\chi \mathcal{T}$. Those types of stars are supposed to be made of strange quark matter (SQM) whose distribution is governed by the phenomenological MIT bag EOS as $p=\frac{1}{3}(\rho-4B)$, where $B$ is the bag constant, while the form of charge distribution is chosen to be $q\left(r\right)=Q\left(r/R\right)^3=\alpha r^3$ with $\alpha$ as a constant. We derive the values of the unknown parameters by matching the interior spacetime to the exterior Reissner-Nordstr{\"o}m metric followed by the appropriate choice of the values of the parameters $\chi$ and $\alpha$. Our study reveals that besides SQM, a new kind of matter distribution originates due to the interaction between the matter and the extra geometric term, while the modification of the Tolman-Oppenheimer-Volkoff (TOV) equation invokes the presence of a new force $F_c$. The accumulation of the electric charge distribution reaches its maximum at the surface, and the predicted values of the corresponding electric charge and electric field are of the order of $10^{19-20}$ C and $10^{21-22}$ V/cm, respectively. To examine the physical validity of our solutions, we perform tests of the energy conditions, stability against equilibrium of the forces, the adiabatic index, etc., and find that the proposed $f(R, \mathcal{T})$ model survives all these critical tests. Therefore, our model can describe the non-singular charged strange stars and justify the supermassive compact stellar objects having their masses beyond the maximum mass limit for the compact stars in the standard scenario. Our model also supports the existence of several exotic astrophysical objects like super-Chandrasekhar white dwarfs, massive pulsars, and even magnetars, which remain unexplained in the framework of General Relativity (GR).}


\begin{document}
\maketitle
\flushbottom

\section{Introduction}\label{sec1}
The cosmological observations of Cosmic Microwave Background Radiation (CMBR)~\cite{Spergel2003,Spergel2007}, high redshift supernovae~\cite{Riess1998}, baryon acoustic oscillations~\cite{Percival2010}, Planck data~\cite{Ade2014}, supernovae of type Ia~\cite{Perlmutter1999,Bennett2003}, etc., lead to the remarkable conclusion that our Universe is not only expanding but also has a phase of the late-time cosmic accelerated expansion. All that immediately questions the fundamental pillar of modern cosmology, i.e., General Relativity (GR). Though at present GR appears to be the best way to study the large-scale structure of the Universe theoretically, the observational evidence of late time acceleration of the Universe becomes a major set back to the acceptance of GR. To explain the observed cosmological dynamics, the standard approach is given by the modification of the Einstein gravitational field equations by the cosmological constant ($\Lambda$), introduced by Einstein~\cite{Sahni2000,Peebles2003}. The Einstein field equations with the cosmological constant provide the best fit to the observed data, with a further assumption of another hypothetical component of the Universe known as {\it Dark Matter}~\cite{Overduin2004,Baer2015}. The present paradigm of the modern age of cosmology is called the $\Lambda$ Cold Dark Matter ($\Lambda$ CDM) model. It is assumed that the Universe is filled with the mysterious component called {\it Dark Energy} which is considered to be the sole reason for the accelerated expansion of the Universe and balances the matter-energy ratio. The dark energy approach seems to be the most suitable resolution of explaining the recent accelerating state of the Universe, though the observed data has a huge discrepancy of the value of 120 orders of magnitude with the theoretically natural value of $\Lambda$~\cite{Weinberg1989,Carroll2001} in quantum gravity.

This situation motivated physicists to develop more sophisticated gravity theories that flourished via extensions of the Einstein-Hilbert action and by using the alternative paradigm known as the modified/extended gravity theories which are different from the standard Einstein gravity description. The modified gravity theories may provide a better understanding of the quantum and gravitational fields at high energy densities. They may also provide the effective approximation to quantum gravity~\cite{Buchbinder1992,Parker2009}. Extending the Einstein-Hilbert Lagrangian density to a function $f\left(R\right)$, where $R$ is the Ricci scalar, allows one to extend GR to the $f(R)$ gravity that attracted much attention~\cite{Capozziello2002,Nojiri2003,Carroll2004,Bertolami2007}. Besides the $f(R)$ gravity, the alternative gravity theories also include the $f\left(G\right)$ gravity~\citep{Bamba2010}, $f\left(R,G\right)$ gravity~\citep{Nojiri2005}, $f\left(\mathbb{T}\right)$ gravity~\citep{Bengochea2009,Linder2010,Bohmer2011}, Brans-Dicke (BD) gravity~\citep{Avilez2014,Bhattacharya2015}, etc., where $R$, $G$ and $\mathbb{T}$ are Ricci scalar, Gauss-Bonnet scalar and the torsion scalar, respectively.

However, the $f(R)$ gravity theory also has difficulties as regards consistency with the Solar system tests~\cite{Erickcek2006,Capozziello2007} and explaining the CMBR tests, the strong lensing regime and the galactic scale~\cite{Dolgov2003,Olmo2005,Yang2007,Dossett2014,Campigotto2017,Tengpeng2018}.  The $f(R)$ gravity theory also fails to justify the existence of a stable stellar configuration~\cite{Briscese2007,Kobayashi2008,Babichev2010}, which raises questions about its validity. These limitations of the $f(R)$ gravity theory forced its further generalizations which include the coupling between the scalar curvature and matter~\cite{Goenner1984,Allemandi2005,Bertolami2007}. In fact, an $f(R)$ gravity is classically equivalent to the scalar-tensor gravity, i.e., the Einstein gravity with a real scalar field having the scalar potential in dual correspondence to the $f$-function~\cite{Fujii2003,Ketov2013}. Hence, 
strictly speaking, the $f(R)$ gravity theories should not be considered as the alternative theories of gravity.  This is also applicable to the higher-dimensional extensions of $f(R)$ gravity~\cite{Nakada2016,Nakada2017} and $F(R)$ supergravity~\cite{Ketov2010,Ketov2011,Ketov2012,Addazi2017}.

Further investigation~\cite{Harko2008,Harko2011} led to a more general approach with arbitrary coupling between matter and geometry, by introducing a new class of generalized gravity models, whose Lagrangian is an arbitrary function of the trace of the energy-momentum tensor ($\mathcal{T}$) and Ricci scalar ($R$), known as the $f(R, \mathcal{T})$ gravity theory. This theory was successfully applied for a description of the late time accelerated expansion of the Universe~\cite{Shabani2014,Baffou2015,Moraes2017}. Barrientos et al.~\cite{Barrientos2018} and Wu et al.~\cite{Wu2018} presented the Palatini approach to the $f(R, \mathcal{T})$ gravity theory by using the connection independent of metric. The generalized form of the hydrostatic stellar equation dubbed as Tolman-Oppenheimer-Volkoff (TOV) equation for the isotropic and anisotropic stellar systems, in the framework of the $f(R, \mathcal{T})$ gravity theory was given in Refs.~\cite{Moraes2016} and~\cite{Deb2018}, respectively. The study by Das et al.~\cite{Das2017}, and Moraes et al.~\cite{Moraes2017} revealed the impact of the $f(R, \mathcal{T})$ gravity in the astrophysical systems like gravastars and wormholes, respectively.
 
 The $f(R, \mathcal{T})$ gravity theory may also provide the effective classical description of some quantum gravity phenomena. For example, Dzhunushaliev et al.~\cite{Dzhunushaliev2014,Dzhunushaliev2015} considered a quantum gravitating system where the specific form of $f(R, \mathcal{T})$ gravity emerges due to quantum fluctuations of metric, when the matter is nonminimally coupled to gravity.  Harko et al.~\cite{Harko2011} made a similar conjecture by recasting quantum effects of gravity into the $\mathcal{T}$ dependence in the matter Lagrangian. This was further explored by considering the formalism of open thermodynamic systems~\cite{Harko2014}. Modified gravity theories related to the curvature-matter coupling exhibit another interesting feature as the non-conservation of their energy-momentum tensor, i.e., ${\nabla}_{\mu} {T}^{\mu\nu} \neq 0$~\citep{Harko2011,Barrientos2014}, which drives particles along non-geodesic paths and, as a result, extra force appears. Thus, the $f(R,\mathcal{T})$ gravity theory violates the equivalence principle (EP) introduced by Einstein to develop GR as the universal principle of physics beyond the Solar system~\cite{Bertolami2008,Damour2010}. By applying the second law of thermodynamics, it is not hard to conclude that the non-conservation of the generalized conservation equation in the $f(R,\mathcal{T})$ gravity theory indicates the irreversible flow of energy from the gravitational field to the newly created matter constituent~\cite{Prigogine1988}. Some studies~\cite{Shabani2017,Josset2017} reveal that the accelerated expansion of the Universe supports violation of the non-conservation of the energy-momentum tensor.
 
Chakraborty~\cite{SC2013} argued that the nonminimal matter-curvature coupling results in a new type of matter originating in the stellar system, while the effective energy-momentum tensor representing a non-interacting two fluid system allows the conservation, i.e., ${\nabla}_{\mu} {T}_{\mu\nu}^{\textit{eff}} = 0$. Further, Deb et al.~\cite{Deb2018b} argued that the extra force due to the $f(R,\mathcal{T})$ gravity is due to the modification of the gravitational Lagrangian, and presented a stable stellar configuration under the equilibrium of the gravitational force, hydrodynamic force and the extra force due to modified gravity. Cosmological and Solar system consequences of the $f(R,\mathcal{T})$ gravity were given by Shabani and Farhaudi~\cite{Shabani2014} where this modified gravity theory was shown to respect the observed data. The consistency of the $f(R,\mathcal{T})$ gravity theory with the gravitational lensing test and the dark matter galactic effects~\cite{Zaregonbadi2016a} also strengthen physical validity of the theory. 

The possible existence of the astrophysical objects entirely made of deconfined up ($u$), down ($d$) and strange ($s$) quarks was suggested in the literature~\cite{Farhi1984,Alcock1986,Haensel1986}, and is known as strange stars. According to the strange quark matter hypothesis~\cite{Bodmer1971,Terazawa1979,Witten1984}, the absolute ground state for the confined state of strongly interacting matter is the strange quark matter (SQM)~\cite{Farhi1984,Alcock1986} rather ${^{56}}Fe$. The observation of massive compact stars with $M\sim 2M_{\odot}$ also advocates the SQM hypothesis. Though it is very hard to differentiate a strange star (SS) from a neutron star (NS), several investigations provided insights to distinguish the compact stellar objects as follows: (i) the SS have higher bulk viscosity~\cite{Haensel1989} and much higher dissipation rate of the radial vibrations~\cite{Wang1984}; (ii) within the first 30 years SS may cool down more rapidly compared to NS~\cite{Schaab1997}; (iii) in comparison to NS the spin rate of SS can be much closer to the Kepler limit~\cite{Madsen1992}, and (iv) the SS has a much lower value of eigenfrequencies of gravitational mode (g-mode) compared to NS~\cite{Fu2008}. Based on the recent observational data acquired by the new generation of X-ray and $\gamma$-ray satellites, the papers~\cite{Li1995,Bombaci1997,Dey1998} demostrated that the compact objects associated with the X-ray burster $4U 1820-30$ and X-ray pulsar $Her X-1$ are consistent with the strange matter equation of state (EOS). Demorest and collaborators~\cite{Demorest2010}, by using Shapiro delay, determined the mass of $PSR~J1614 + 2230$ and found that such highly massive stellar system is consistent with the SQM EOS. Gangopadhyay et al.~\cite{Gangopadhyay2013} predicted radii of the 12 possible candidates of SS.

The SS made of approximately equal numbers of $u$, $d$ and $s$ quarks also includes a smaller number of electrons to maintain global charge neutrality. The investigations by Alcock et al.~\cite{Alcock1986,Alcock1988} and Usov et al.~\cite{Usov2004,Usov2005} reveal that the surface of SS has a high electric field of the order of ${10}^{18-19}$~V~{cm}$^{-1}$, while the presence of electrons on the surface triggers a layer of electric dipoles there. It should be noticed that the energy density associated with such tremendously high value of the electric field should be included into the energy-momentum tensor, because their values are of the same order as the energy density of SQM. The non-zero charge affects the relativistic stellar system as follows: (i) the high electric field affects the spacetime metric, (ii) it modifies the relativistic hydrostatic equilibrium equation with the Coulomb interaction, and (iii) the energy density related to the electric field contributes to the gravitational mass of the stellar system. The effects of electric charge on the spherically symmetric compact stellar system were extensively studied in the literature, see \cite{Ray2008,Negreiros2009,Varela2010,Rahaman2012,Deb2018c} and the references therein.

In this paper, we consider the specific SQM and the electric charge distribution in the particular SS model in the framework of $f(R,\mathcal{T})$ gravity. The outline of our study is as follows: the background behind the choice of the MIT bag EOS and the form of the electric charge distribution are discussed in Sec.~\ref{sec2}. The formalism of the $f(R,\mathcal{T})$ gravity theory is given in Sec.~\ref{sec3},  and the essential stellar structure equations are introduced in Sec.~\ref{sec4}. The solutions to the Einstein-Maxwell field equations and the physical parameters are derived in Sec.~\ref{sec5}. We investigate physical validity of the $f(R,\mathcal{T})$ gravity model via different tests in Sec.~\ref{sec6}. We conclude with a detailed discussion on the achieved results and their implications on the various astrophysical situations in Sec.~\ref{sec7}.

\section{MIT bag EOS and electric charge distribution}\label{sec2}

To explain hadrons, Chodos et al.~\cite{Chodos1974} proposed the phenomenological MIT bag model EOS. It was assumed that the asymptotically free quarks are confined in a finite region of space (say, inside a "bag"). The bag constant $B$ defined as a constant energy per unit volume of the space, may also be considered  as the inward pressure required to confine quarks inside the bag. In our investigation, we assume that the SQM distribution is governed by the MIT bag model EOS. For simplicity, it is assumed that $u$, $d$ and $s$ quarks are non-interacting and massless. Thus, according to the MIT bag model, the quark pressure $p$ is defined as
\begin{equation}\label{2.1}
{p}={\sum_{f=u,d,s}}{p^f}-{B},
 \end{equation}
where $p^f$ is the pressure due to each flavor. The energy density of each flavor ${\rho}^f$ is related to the corresponding pressure $p^f$ by the relation $p^f=\frac{1}{3}{{\rho}^f}$. 

The energy density due to the SQM distribution in MIT bag model is defined as
\begin{equation}
{{\rho}}={\sum_{f=u,d,s}}{{\rho}^f}+B~. \label{2.2}
\end{equation}
By using Eqs.~\eqref{2.1} and \eqref{2.2} and the relation between ${\rho}^f$ and $p^f$, we
arrive at  the well-known simplified MIT bag model EOS as follows:
 \begin{equation}
{p}=\frac{1}{3}({{\rho}}-4\,B).\label{2.3}
\end{equation}

This MIT bag model EOS was used to discuss the stellar systems made of SQM distribution in the  literature, see \cite{Negreiros2009,Rahaman2012,Brilenkov2013,Paulucci2014,Deb2017} and references therein. To derive the numerical output of the model, we performed all our calculations with $B$=83~MeV {fm}$^{-3}$~\cite{Rahaman2014} as the reference value of $B$, which is well within the possible range provided by other authors~\cite{Farhi1984,Alcock1986,Burgio2002}. 

As regards the effects of electric charge on the strange star configuration, we follow de Felice et al.~\cite{Felice1995} and consider the radial coordinate dependent form of the electric charge $q(r)$ for the parametric value of $n=3$ as follows:
\begin{eqnarray}\label{2.4}
q\left(r\right)=Q\left( \frac{r}{R} \right)^3\equiv \alpha\,{r^3},
\end{eqnarray}
where $Q$ and $R$ are the total charge and radius of the stellar system, and  $\alpha$ is a constant. We take $\alpha$=0.0010~km$^{-2}$ as the reference value.

\section{$f\left(R,\mathcal{T}\right)$ gravity in Einstein-Maxwell spacetime}\label{sec3}

Following the work of Harko et al.~\cite{Harko2011}, we modify the standard form of the gravitational Lagrangian by replacing an arbitrary function of Ricci scalar $R$ and trace of the energy momentum tensor $\mathcal{T}$, i.e., $f(R, \mathcal{T})$ and define the modified from of the Einstein-Hilbert action in the Einstein-Maxwell space as follows:
\begin{eqnarray}\label{3.1}
S=\frac{1}{16\pi}\int d^{4}xf(R,\mathcal{T})\sqrt{-g}+\int d^{4}x\mathcal{L}_m\sqrt{-g}+\int d^{4}x\mathcal{L}_e\sqrt{-g},
\end{eqnarray}
where $T_{\mu\nu}$ is the energy-momentum tensor of the SQM distribution, ${\mathcal{L}}_m$ represents the Lagrangian for the matter distribution and ${\mathcal{L}}_e$ denotes the Lagrangian for the electromagnetic field. 

We define $T_{\mu\nu}$ as follows
\begin{eqnarray}\label{3.2}
T_{\mu\nu}=-\frac{2}{\sqrt{-g}}\frac{\delta(\sqrt{-g} {{\mathcal{L}}_m})}{\delta g^{\mu\nu}},
\end{eqnarray}
and also define the trace of $T_{\mu\nu}$ as $\mathcal{T}=g^{\mu\nu}T_{\mu\nu}$. As ${\mathcal{L}}_m$ depends only on the metric tensor components $g_{\mu\nu}$ and not on their derivatives, so we find
\begin{eqnarray}\label{3.3}
T_{\mu\nu}=g_{\mu\nu} {\mathcal{L}}_m - 2 \frac{\partial {\mathcal{L}}_m}{\partial {g^{\mu\nu}}}.
\end{eqnarray}

Again, ${\mathcal{L}}_e$ in Eq.~\eqref{3.1} represents Lagrangian of the electromagnetic field defined as
\begin{eqnarray}\label{3.4}
{\mathcal{L}}_e=-\frac{1}{16\pi}F_{\alpha\beta} F_{\gamma\sigma} g^{\alpha\gamma} g^{\beta\sigma},
\end{eqnarray}
where $F_{\alpha\beta}$ is the electromagnetic field tensor and defined in terms of electromagnetic four potential $A_{\alpha}$ as $F_{\alpha\beta}={\partial}_{\alpha} A_{\beta} - {\partial}_{\beta} A_{\alpha}$.

Now, varying the action~\eqref{3.1} with respect to the metric tensor component $g_{\mu\nu}$ we obtain the field equations of the model in $f(R,\mathcal{T})$ gravity theory as follows:
\begin{eqnarray}\label{3.5}
& & \hspace{-1cm} G_{\mu\nu}=\frac{1}{f_R(R,\mathcal{T})}\Bigg[8\pi T_{\mu\nu}+\frac{1}{2} f(R,\mathcal{T}) g_{\mu\nu}-\frac{1}{2}R{f_R(R,\mathcal{T})}g_{\mu\nu} - (T_{\mu\nu}+\Theta_{\mu\nu}){f_T(R,\mathcal{T})}\nonumber \\
& & \hspace{5cm} +\left(\nabla_{\mu} \nabla_{\nu}-g_{\mu\nu}\DAlambert\right) {{f_R}\left(R,\mathcal{T}\right)}+8\pi E_{\mu\nu}\Bigg],
\end{eqnarray}
where we define ${f_R (R,\mathcal{T})= \frac{\partial f(R,\mathcal{T})}{\partial R}}$, $\Theta_{\mu\nu}=\frac{g^{\alpha\beta}\delta T_{\alpha\beta}}{\delta g^{\mu\nu}}$ and ${f_\mathcal{T}(R,\mathcal{T})=\frac{\partial f(R,\mathcal{T})}{\partial \mathcal{T}}}$. Here $\DAlambert \equiv\partial_{\mu}(\sqrt{-g}\\
 g^{\mu\nu} \partial_{\nu})/\sqrt{-g}$ is the D'Alambert operator, $R_{\mu\nu}$ is the Ricci tensor, $\nabla_\mu$ represents the covariant derivative associated with the Levi-Civita connection of $g_{\mu\nu}$, $G_{\mu\nu}$ is the Einstein tensor and $E_{\mu\nu}$ is the electromagnetic energy-momentum tensor. 

To represent charged isotropic SQM distribution, $T_{\mu\nu}$ and $E_{\mu\nu}$ can be defined as
\begin{eqnarray}\label{3.6}
& & T_{\mu\nu}=\left( \rho+p \right)u_{\mu} u_{\nu}+p g_{\mu\nu}, \\ \label{3.7}
& & E_{\mu\nu}=\frac{1}{4\pi}\left(F^{\gamma}_{\mu} F_{\nu \gamma}-\frac{1}{4} g_{\mu \nu} F_{\gamma \beta} F^{\gamma \beta} \right),
\end{eqnarray}
where $u_{\mu}$ is the four velocity which satisfies the conditions $u_{\mu}u^{\mu}=1$ and $u^{\mu} {\nabla}_{\nu} u_{\mu}=0$, respectively, $\rho$ and $p$ represent SQM density and pressure, respectively. In the present work, we consider $\mathcal{L}_m=-p$ and we obtain $\Theta_{\mu\nu}=-2 T_{\mu\nu} - p g_{\mu\nu}$. 

Now, the covariant divergence of Eq.~\eqref{3.5} reads~\cite{Barrientos2014}
\begin{eqnarray} \label{3.8}
& & \hspace{-1cm} {\nabla}^{\mu} T_{\mu\nu}=\frac{f_T \left(R,\mathcal{T}\right)}{8\pi-f_T \left(R,\mathcal{T}\right)}\Big[\left(T_{\mu\nu}+\Theta_{\mu\nu} \right){\nabla}^{\mu} lnf_T \left(R,\mathcal{T}\right)+{{\nabla}^{\mu}}\Theta_{\mu\nu} -\frac{1}{2}g_{\mu\nu}{\nabla}^{\mu} T \nonumber \\ 
& & \hspace{9cm} -\frac{8\pi}{f_T \left(R,\mathcal{T}\right)}{\nabla}^{\mu} E_{\mu\nu}\Big], 
\end{eqnarray}
which reveals that likewise the other extended gravity theories Eq.~\eqref{3.8} also features non-conservation of the energy-momentum tensor due to SQM distribution~\cite{Zhao2012,Yu2018}.

In the present article, we consider the simplest linear form of the function $f(R, \mathcal{T})$ as $f(R, \mathcal{T})=R+2\chi \mathcal{T}$, where $\chi$ is the matter-geometry coupling constant~\cite{Harko2011}. Note that using this specific form of $f(R, \mathcal{T})$ function many researchers~\cite{Shamir2015,Moraes2015a,Deb2018,Deb2018d} have successfully studied $f(R, \mathcal{T})$ gravity theory. Substituting this specific form of $f(R, \mathcal{T})$ function in Eq.~\eqref{3.5} the field equation for $f(R, \mathcal{T})$ gravity theory reads
\begin{eqnarray} \label{3.9}
& & \hspace{0cm} G_{\mu\nu} = (8\pi+2\chi)T_{\mu\nu}+2\chi p g_{\mu\nu}+\chi \mathcal{T} g_{\mu\nu}+8\pi E_{\mu\nu}  =8\pi \left(T^{\textit{eff}}_{\mu\nu}+E_{\mu\nu}\right)=8\pi T_{ab},
\end{eqnarray}
where $T_{ab}=T^{\textit{\textit{eff}}}_{\mu\nu}+E_{\mu\nu}$ represents the energy-momentum tensor of the charged effective matter distribution and $T^{\textit{\textit{eff}}}_{\mu\nu}$ represents energy-momentum tensor of the effective fluid, i.e., SQM and the new kind of fluid which originates due to the matter geometry coupling, given as
\begin{eqnarray} \label{3.10}
T^{\textit{eff}}_{\mu\nu}=T_{\mu\nu}\left(1+\frac{\chi}{4\pi}\right)+\frac{\chi}{8\pi}\left(\mathcal{T}+2p\right) g_{\mu\nu}.
\end{eqnarray}

Substituting $f(R, \mathcal{T})=R+2\chi \mathcal{T}$ in Eq.~\eqref{3.8}, we obtain
\begin{eqnarray} \label{3.11}
& & \hspace{-0.2cm}  \left(4\pi+\chi \right) {\nabla}^{\mu} T_{\mu\nu}= -\frac{1}{2} \chi \Big[g_{\mu\nu} {\nabla}^{\mu}\mathcal{T}+2 {\nabla}^{\mu}\left(p g_{\mu\nu} \right)+\frac{8\pi}{\chi} E_{\mu\nu} \Big].
\end{eqnarray}

However, Eq.~\eqref{3.11} can be written in the following workable form:
\begin{eqnarray} \label{3.12}
& & {\nabla}^{\mu} T_{ab}=0.
\end{eqnarray}

Eq.~\eqref{3.12} represents conservation of the energy-momentum tensor $T^{\textit{\textit{eff}}}_{\mu\nu}$ for the effective matter distribution. Note that for $\chi=0$ the standard results due to GR can be achieved.
Of course, the matter-geometry coupling term $2\chi\mathcal{T}$ in the full action~\eqref{3.1} can also be included in the matter Lagrangian instead of modifying the gravitational one. This would lead to a modification of the matter couplings in the context of GR. Both the descriptions are equivalent and hence lead to the same physics. The existence of this correspondence is a great simplification of the linear ansatz $f(R,\mathcal{T})=R + 2\chi\mathcal{T}$ versus more general and much more complicated $f(R,\mathcal{T})$ gravity theories. In our paper, we use the modified gravity approach.

\section{Basic stellar equations in $f(R, \mathcal{T})$ gravity} \label{sec4}
To describe interior spacetime of the spherically symmetric static stellar system, we take metric as follows:
\begin{eqnarray}\label{3.13}
ds^2=e^{\nu(r)}dt^2-e^{\lambda(r)}dr^2-r^2(d\theta^2+\sin^2\theta d\phi^2),
\end{eqnarray} 
where the metric potentials $\nu$ and $\lambda$ are the functions of the radial coordinate $r$ only.

The electromagnetic field tensor $F_{\alpha\beta}$ satisfies the covariant Maxwell equations
\begin{eqnarray}\label{3.14}
F_{[\alpha\beta,\gamma]}=0,  \\ \label{3.15}
{F^{\mu\nu}}_{;\nu}=4\pi j^{\mu},
\end{eqnarray}
where $j^{\mu}$ is the electromagnetic 4-current vector
\begin{eqnarray}\label{3.16}
j^{\mu} = \frac{\sigma}{\sqrt {g_{44}}} \frac{dx^i}{dx^4} = \sigma v^i,
\end{eqnarray}
where $\sigma=e^{\nu /2} J^{0}\left(r\right)$ denotes the charge density.

In the spherically symmetric static stellar system the only non vanishing component of $J^{\mu}$ is given as $J^0$ and the only non-vanishing components of the Maxwell tensor are 
\begin{eqnarray}\label{3.17}
F^{01}=-F^{10}=\frac{q(r)}{r^2} e^{-\left(\nu+\lambda\right)/2},
\end{eqnarray}
where $q(r)$ is the charge enclosed by a spherical stellar system of radius $r$. It  can be defined by considering the relativistic form of the Gauss law as follows:
\begin{eqnarray}\label{3.18}
q(r) = 4\pi \int_0^r \sigma r^{\prime 2} e^{\lambda/2} dr^{\prime} = r^2 \sqrt{-F_{10}F^{10}}~. 
\end{eqnarray}
It follows that
\begin{eqnarray}\label{3.19}
E^2=-F^{\alpha\beta}F_{\alpha\beta}=\frac{q^2}{r^4},
\end{eqnarray}
where $E$ represents the electric field. 

Substituting Eqs. \eqref{3.6} and \eqref{3.7} into Eq. \eqref{3.9} we find the explicit form of the Einstein field equation for the interior metric \eqref{3.13} as follows:
\begin{eqnarray}\label{3.21}
& \hspace{-1cm} {{\rm e}^{-\lambda }} \left( {\frac {\lambda^{{\prime}}}{r}}-\frac{1}{r^2}\right) +\frac{1}{r^2}= \left(8\pi+3\chi\right)\rho-\chi p+\frac{q^2}{r^4}  =8\pi \rho^{\textit{eff}}+\frac{q^2}{r^4}, \\ \label{3.22}
& \hspace{-1cm} {{\rm e}^{-\lambda}} \left( {\frac {\nu^{{\prime}}}{r}}+\frac{1}{r^2}\right) -\frac{1}{r^2}=\left(8\pi+3\chi\right)p-\chi\rho-\frac{q^2}{r^4}  =8\pi  p^{\textit{eff}}-\frac{q^2}{r^4},
\end{eqnarray}
where the primes denote  the differentiation with respect to the radial coordinate $r$. Here $\rho^{\textit{eff}}$ and $p^{\textit{eff}}$ represent the effective density and pressure of the matter distribution, respectively, 
\begin{eqnarray}\label{3.23}
& \rho^{\textit{eff}}=\rho+\frac{\chi}{8\pi}\left(3\rho-p\right), \\ \label{3.23a}
& p^{\textit{eff}}=p-\frac{\chi}{8\pi}\left(\rho-3p\right).
\end{eqnarray}

The essential stellar structure equations required to describe static and charged spherically symmetric sphere in $f(R, \mathcal{T})$ gravity theory are given by
\begin{eqnarray}\label{3.24}
& \frac{dm}{dr}=4\pi \rho r^2+\frac{q}{r} \frac{dq}{dr}+\frac{\chi}{2}\left(3\rho-p\right)r^2, \\ \label{3.25}
&\hspace{-0.2cm} \frac{dp}{dr}=\frac{1}{\left[1+\frac{\chi}{8\pi+2\chi}\left(1-\frac{d\rho}{dp}\right)\right]}\Bigg\lbrace -\left(\rho+p\right)\bigg[\Big\lbrace 4\pi\rho r+\frac{m}{r^2} -\frac{q^2}{r^3}  -\frac{\chi}{2}\left(\rho-3p\right)r \Big\rbrace \bigg/ \left(1-\frac{2m}{r}+\frac{q^2}{r^2}\right)\bigg]\nonumber \\
& \hspace{4cm} + \frac{8\pi}{8\pi+2\chi}\frac{q}{4\pi r^4}\frac{dq}{dr} \Bigg\rbrace, \nonumber \\
\end{eqnarray}  
where the metric potential $e^{\lambda}$ has the Reisner-Nordstr{\"o}m-type form
\begin{equation}
e^{\lambda}=1-\frac{2m}{r}+\frac{q^2}{r^2}.
\end{equation}

Following Mak and Harko~\cite{Mak2002}, for a non-singular and monotonically decreasing SQM density function inside the spherically symmetric stellar system, we consider the simplified form 
of $\rho$ as
\begin{equation} \label{3.26}
\rho=\rho_c \left[1-\left(1-\frac{\rho_0}{\rho_c}\right) \frac{r^2}{R^2} \right],
\end{equation}
where constant $\rho_c$ and $\rho_0$ are values of $\rho$ at the centre and on the surface of the spherical stellar system, respectively.

Outside the stellar system $\mathcal{T}=0$, due to the absence of the charged matter distribution  the standard form of the exterior Reissner-Nordstr{\"o}m metric is still applicable to describe the exterior spacetime as follows:
\begin{eqnarray}\label{3.27}
&\qquad\hspace{-1cm} ds^2 = \left(1 - \frac{2M}{r} +\frac{Q^2}{r^2}\right) dt^2- \frac{1}{\left(1 - \frac{2M}{r} + \frac{Q^2}{r^2}\right)} dr^2 - r^2(d\theta^2 + \sin^2\theta d\phi^2).
\end{eqnarray}

\section{Solution to the Einstein-Maxwell field equation in $f\left(R,\mathcal{T}\right)$ gravity}\label{sec5}

Substituting Eqs.~\eqref{2.3}, \eqref{2.4}, \eqref{3.24}, \eqref{3.26} and \eqref{3.27} into Eqs. \eqref{3.21} and \eqref{3.22} we obtain the expressions for the physical parameters, viz., $\lambda$, $\nu$, $\rho^{\textit{eff}}$ and $p^{\textit{eff}}$ as follows: 
\begin{eqnarray} \label{3.28}
& \hspace{-2cm}  \lambda=-\ln  \Big[ \big\lbrace 3\,{R}^{7}{r}^{2}\pi \,{\alpha}^{2}+ ( -2\,{\alpha}^{2}{r}^{4}\pi +16\,{r}^{2}\nu_{{1}} +\pi ) {R}^{5} -16\,{r}^{4}\nu_{{1}}{R}^{3}-5\,M{R}^{2}{r}^{2}\pi \nonumber \\
& \hspace{2cm} +3\,M{r}^{4} \pi \big\rbrace \bigg/ {\pi \,{R}^{5}} \Big], \\ \label{3.29}
& \hspace{-2cm} \nu=-\ln  \big\lbrace 3\,{R}^{7}{r}^{2}\pi \,{\alpha}^{2}+ \left( -2\,\pi \,{
\alpha}^{2}{r}^{4}+16\,{r}^{2}\nu_{{1}}+\pi  \right) {R}^{5} -16\,{r}^{4}\nu_{{1}}{R}^{3}-5\,M{R}^{2}{r}^{2}\pi \nonumber \\ 
& +3\,M{r}^{4}\pi  \big\rbrace \nu_{{3}}\nu_{{5}} +\nu_{{4}}\nu_{{5}}{\rm arctanh} \Big\lbrace{\frac {{R}^{2} \left( \pi \,{R}^{5}{\alpha}^{2}+16\nu_{{1}}{R}^{3}-\pi \,M \right) }{16\nu_{{2}}}}\Big\rbrace  + \nu_{{3}}\nu_{{5}}\ln  \Big\lbrace -{R}^{4}\pi \, \big( -{R}^{5}{\alpha}^{2} \nonumber \\ 
& +2\,M-R\big)\Big\rbrace +\nu_{{4}}\nu_{{5}} {\rm arctanh} \bigg[\Big\lbrace \big( 3\,{R}^{7}{\alpha}^{2}-4\,{R}^{5}{\alpha}^{2}{r}^{2}-5\,M{R}^{2}+6\,M{r}^{2} \big) \pi  \nonumber \\ 
& +16\,{R}^{3}\nu_{{1}} \left( {R}^{2}-2\,{r}^{2} \right)\Big\rbrace \bigg/ {16 \nu_{{2}}}\bigg]+\ln  \left( 1-{\frac {2M}{R}}+{\alpha}^{2}{R}^{4} \right),  \\ \label{3.30}
& {\rho}^{{\it eff}}= \Big[\big\lbrace -9\,{R}^{7}\pi \,{\alpha}^{2}+ [-48\,B{\pi }^{2}+ ( 9\,{\alpha}^{2}{r}^{2} -36\,B\chi ) \pi -6\,B{\chi}^{2} ] {R}^{5}+80\,{r}^{2}\nu_{{1}}{R}^{3}+15\,M{R}^{2}\pi \nonumber \\
& -15\,M{r}^{2}\pi \big)\big\rbrace \bigg/ 8{R}^{5}{\pi }^{2} \Big],   \\ \label{3.31}
& \hspace{-8cm} {p}^{{\it eff}}=-\frac{10\, \left( {\frac {9}{80}}\pi {R}^{5}{\alpha}^{2}+\nu_{{1}}{R}^{3
}-\frac{3}{16}\,\pi \,M \right)  \left( {R}^{2}-{r}^{2} \right)}{{R}^{5}\pi \, \left( 3\,\pi +\chi \right) },
\end{eqnarray}
where the values of the constants $\nu_1$, $\nu_2$, $\nu_3$, $\nu_4$ and $\nu_5$ are given in APPENDIX.

%%%%%%%%%%%%%%%%%%%%%%%%%%%%%%%%%%%%%%%%%%%%%%%%%%%%%%%%%%%%%%
\begin{figure}
\centering
    \subfloat{\includegraphics[width=6cm]{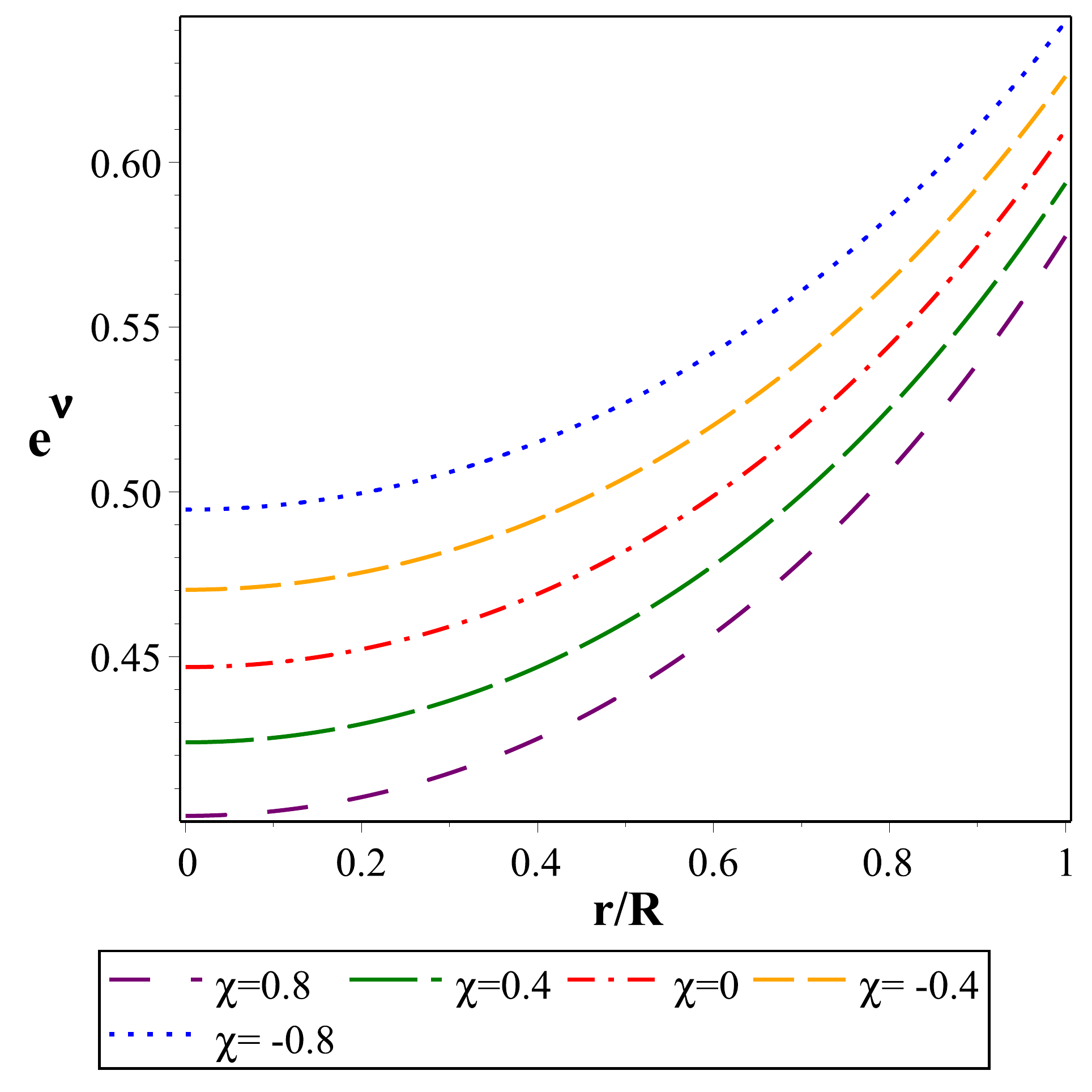}}
    \
    \subfloat{\includegraphics[width=6cm]{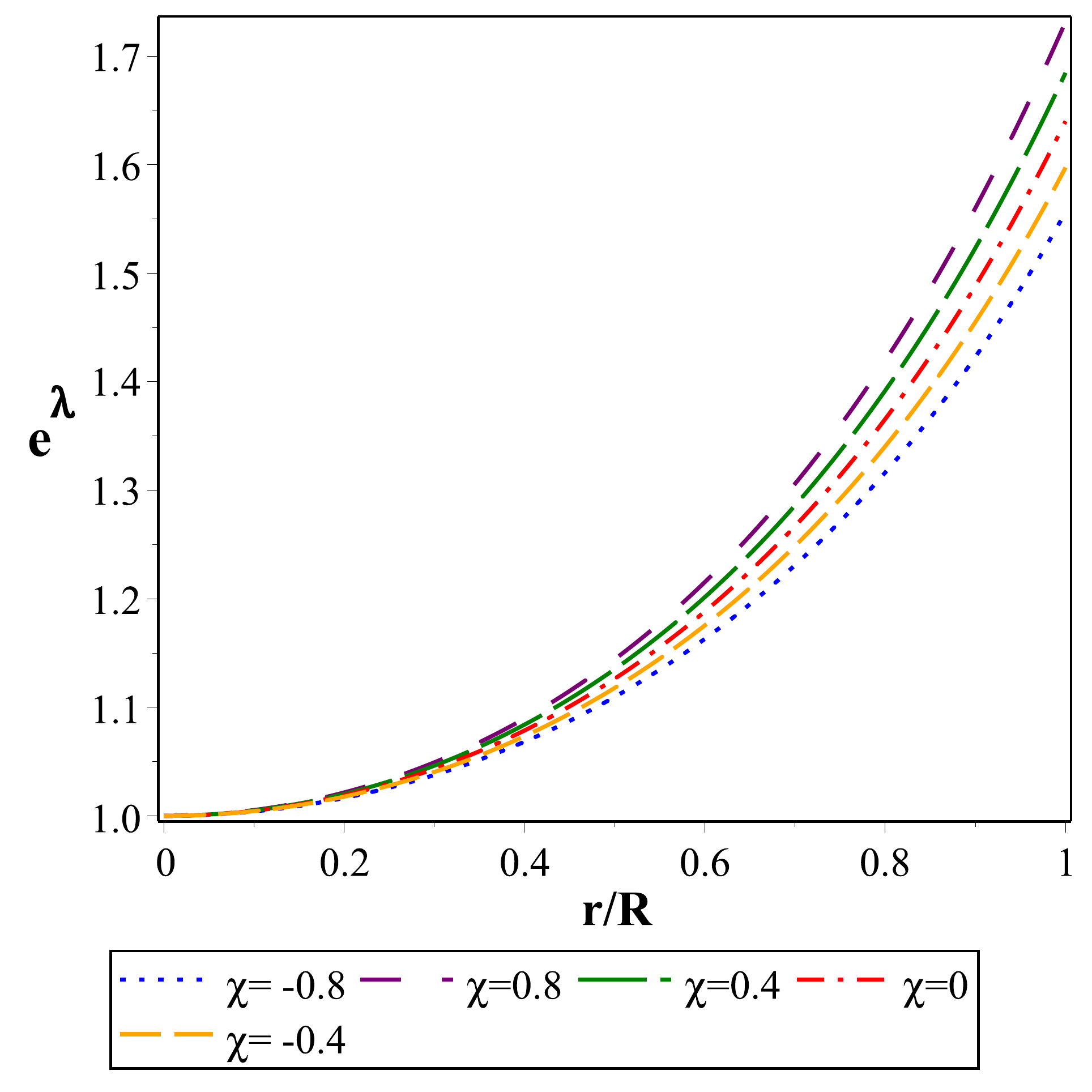}}
    \caption{The metric potentials $e^{\nu}$ and $e^{\lambda}$ as the functions of the radial coordinate $r/R$ for $LMC\,X-4$.} \label{Fig1}
\end{figure}
%%%%%%%%%%%%%%%%%%%%%%%%%%%%%%%%%%%%%%%%%%%%%%%%%%%%%%%%%%%%%%   

%%%%%%%%%%%%%%%%%%%%%%%%%%%%%%%%%%%%%%%%%%%%%%%%%%%%%%%%%%%%%%
\begin{figure}
\centering
    \subfloat{\includegraphics[width=6cm]{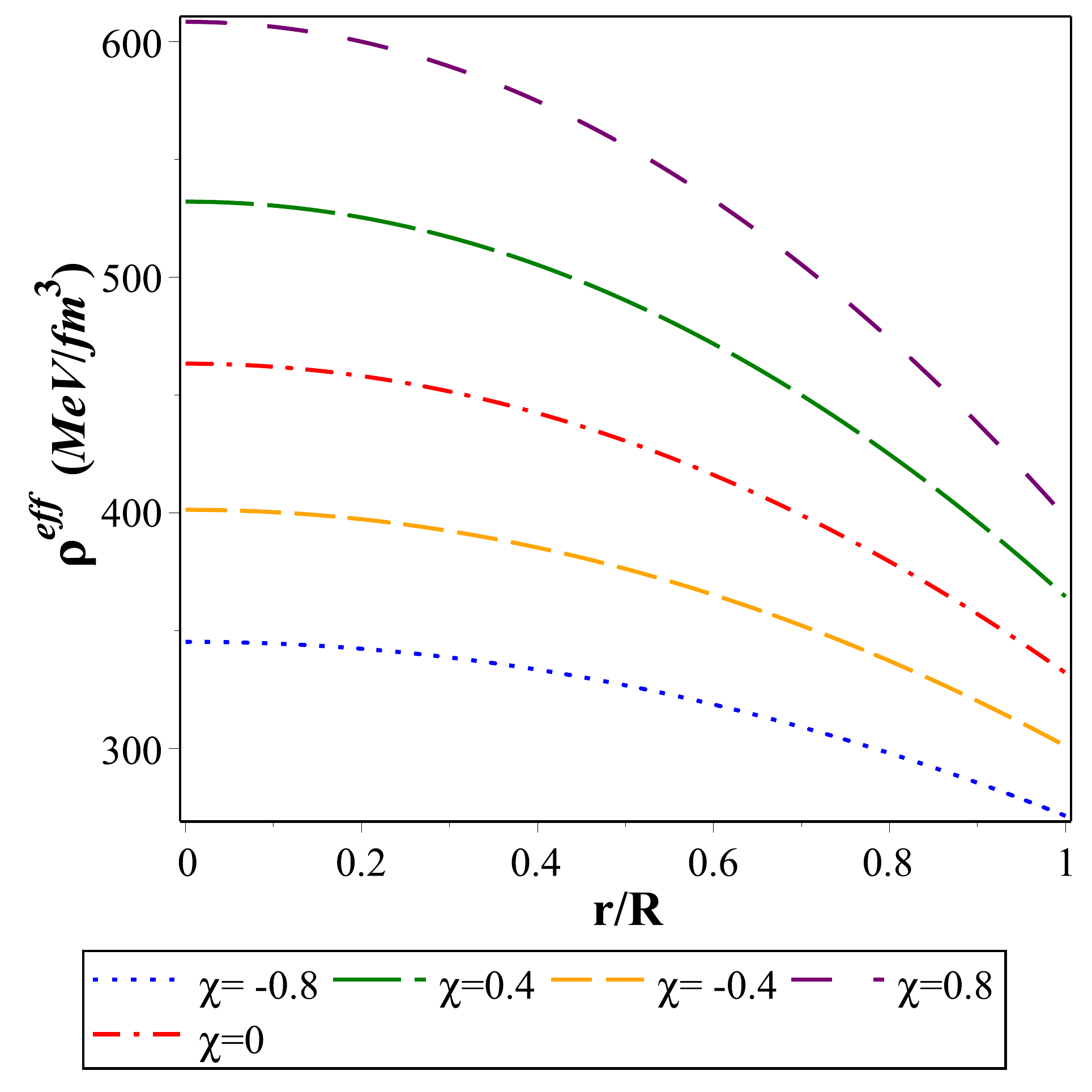}}
    \
    \subfloat{\includegraphics[width=6cm]{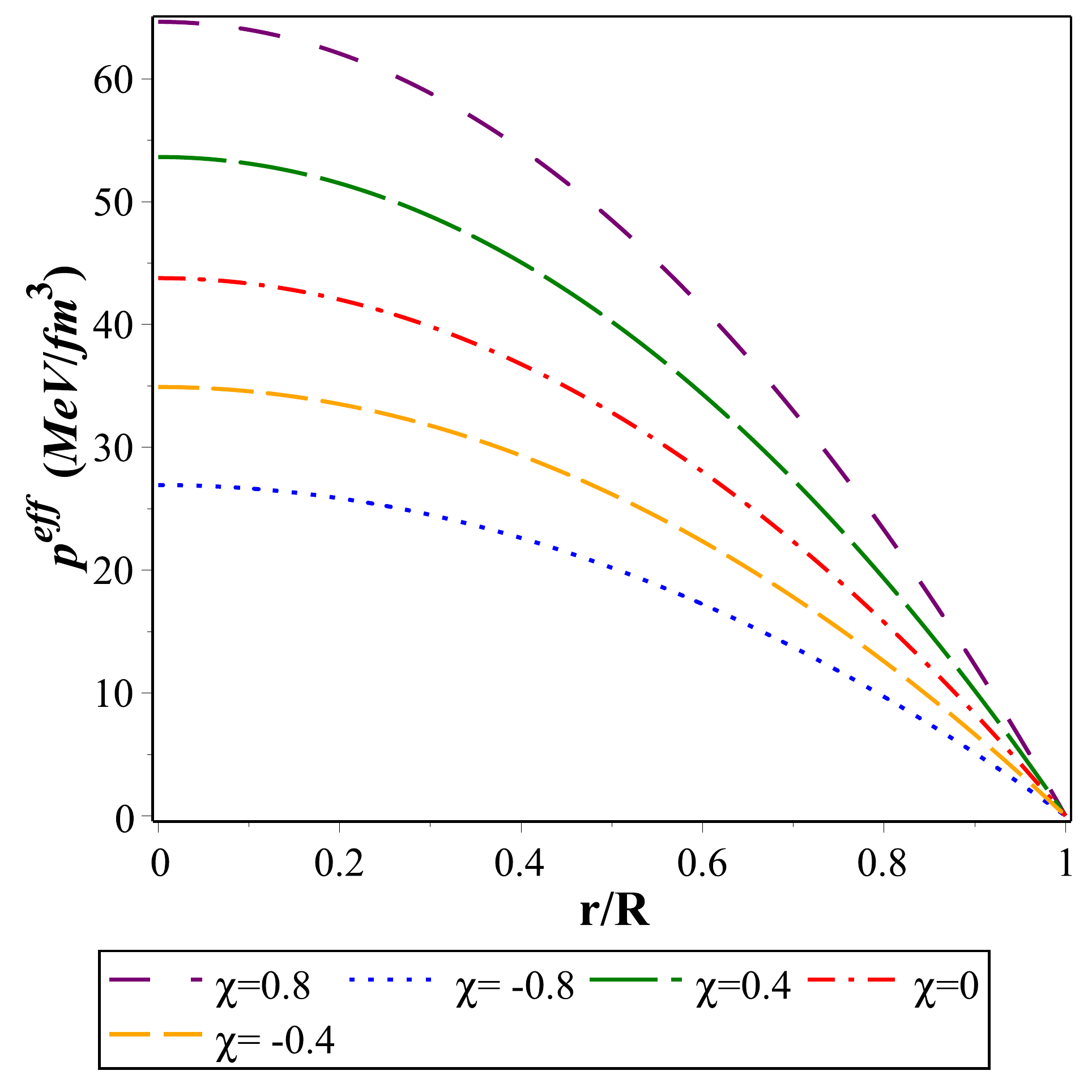}}
    \caption{The effective density and pressure as the functions of the radial coordinate $r/R$ for $LMC\,X-4$.} \label{Fig2}
\end{figure}
%%%%%%%%%%%%%%%%%%%%%%%%%%%%%%%%%%%%%%%%%%%%%%%%%%%%%%%%%%%%%%  

%%%%%%%%%%%%%%%%%%%%%%%%%%%%%%%%%%%%%%%%%%%%%%%%%%%%%%%%%%%%%%
\begin{figure}
\centering
    \subfloat{\includegraphics[width=6cm]{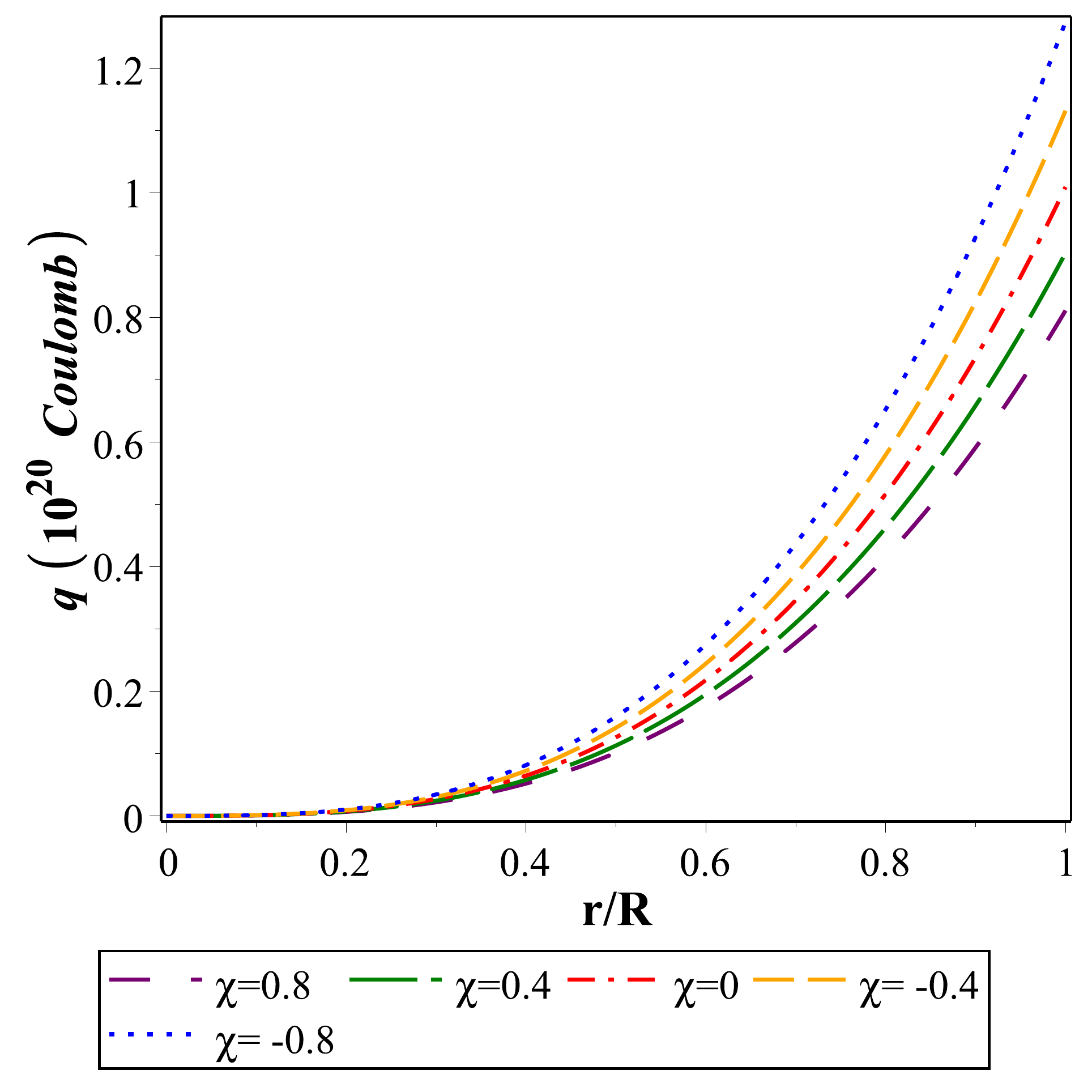}}
    \
    \subfloat{\includegraphics[width=6cm]{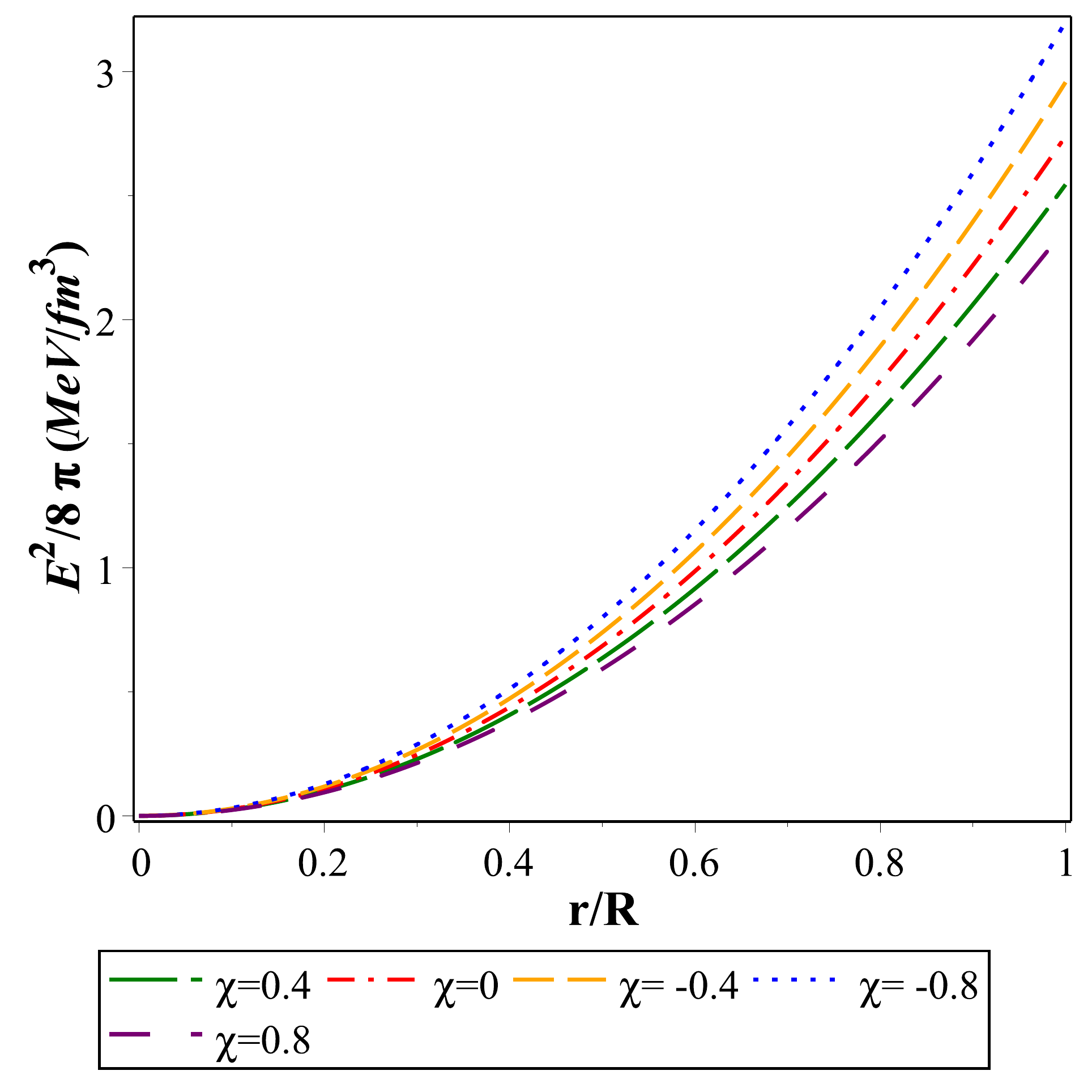}}
    \caption{The electric charge  and the electric energy density as the functions of the radial coordinate $r/R$ for $LMC\,X-4$.} \label{Fig3}
\end{figure}
%%%%%%%%%%%%%%%%%%%%%%%%%%%%%%%%%%%%%%%%%%%%%%%%%%%%%%%%%%%%%%  

We feature the variation of the metric potentials $e^{\nu}$ and $e^{\lambda}$ with respect to the fractional radial coordinate $r/R$ on the left and right panels of Figure~\ref{Fig1}, respectively. The behaviour of the effective energy density $\rho^{\textit{eff}}$ and the effective pressure $p^{\textit{eff}}$ with respect to $r/R$ are shown on the left and right panels of figure~\ref{Fig2}, respectively, which monotonically decrease from the centre to reach their minimum value at the surface. The variation of the electric charge function $q(r)$ and electrical energy density $E^2(r)/8\pi$ with respect to $r/R$ are shown on the left and right panels of figure~\ref{Fig3}, which show that both $q(r)$ and $E^2(r)/8\pi$ reach their minimal values equal to zero at the centre and monotonically increase to reach their maximal values at the surface. Therefore, our solution set 
shows the physically viable and non-singular character of the $f\left(R,\mathcal{T}\right)$ gravity
model under investigation.

\section{Physical features of charged quark stars in $f\left(R,\mathcal{T}\right)$ gravity}\label{sec6}
In this section, we study some physical features of the charged compact star, in order to examine the physical validity and stability of the system  in the $f\left(R,\mathcal{T}\right)$ gravity.

\subsection{Energy conditions}\label{subsec6.1}

Based on the obtained field equation~\eqref{3.9}, the corresponding energy conditions, viz., the Null Energy Condition (NEC), the Weak Energy Condition (WEC), the Strong Energy Condition (SEC) and the Dominant Energy Condition (DEC) in $f\left(R,\mathcal{T}\right)$ gravity theory are given by
{{\begin{eqnarray}\label{6.1.1}
& & ~NEC:~~{{\rho}^{\textit{eff}}}+{{p}^{\textit{eff}}}\geq 0,\\ \label{6.1.2}
& & ~WEC:~~{{\rho}^{\textit{eff}}}+{{p}^{\textit{eff}}}\geq 0,~{{\rho}^{\textit{eff}}+\frac{E^2}{8\pi}}\geq 0, \\ \label{6.1.3}
& & ~SEC:~~{{\rho}^{\textit{eff}}}+{{p}^{\textit{eff}}}\geq 0,~{{\rho}^{\textit{eff}}+3{{p}^{\textit{eff}}}+\frac{E^2}{4\pi}}, \\ \label{6.1.4}
& &   ~DEC:~~{{\rho}^{\textit{eff}}}+\frac{E^2}{8\pi}\geq 0,~{{\rho}^{\textit{eff}}-{{p}^{\textit{eff}}}+\frac{E^2}{4\pi}}\geq 0.
 \end{eqnarray}}}

%%%%%%%%%%%%%%%%%%%%%%%%%%%%%%%%%%%%%%%%%%%%%%%%%%%%%%%%%%%%%%
\begin{figure}
\centering
    \includegraphics[width=6cm]{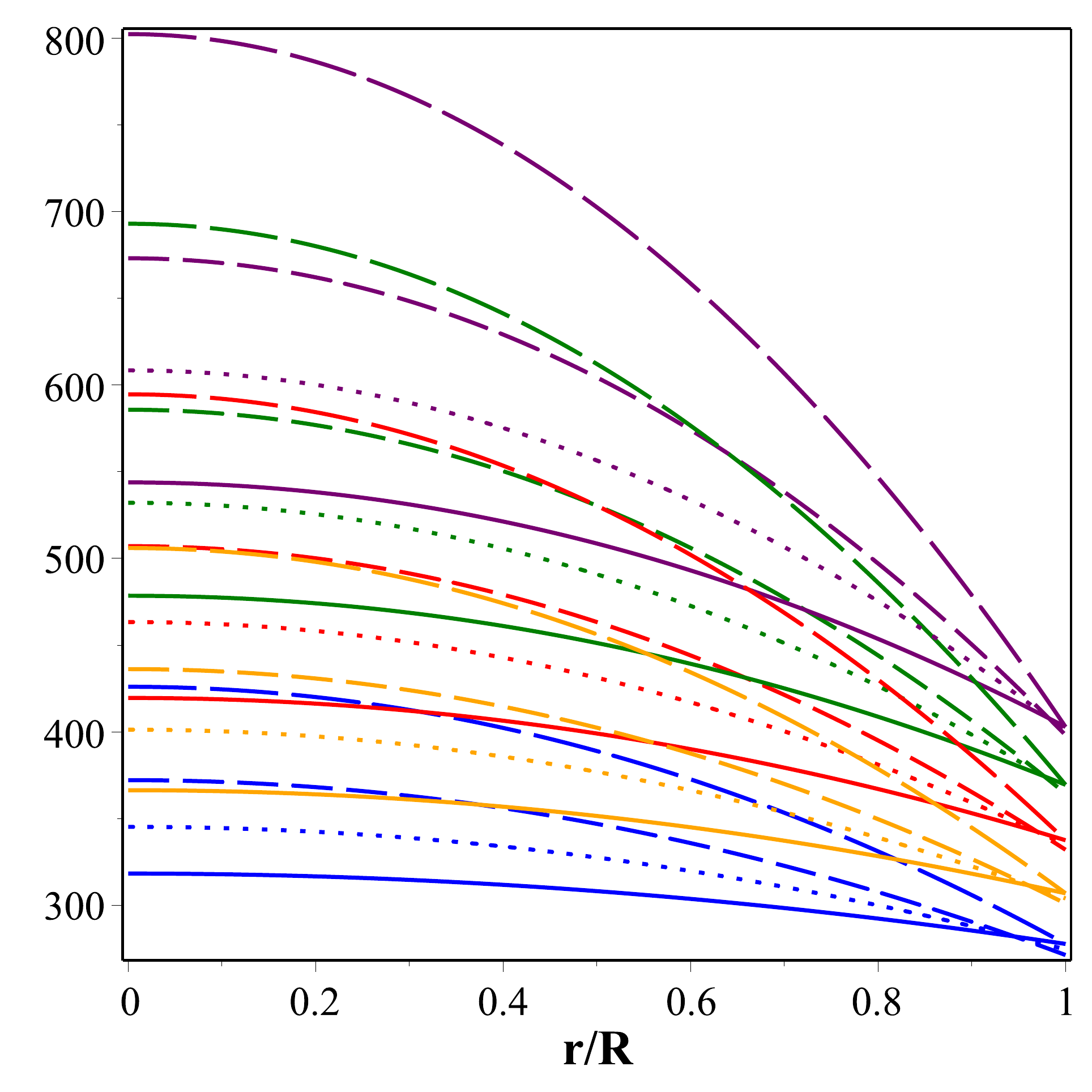}
    \caption{The energy conditions as the functions of the radial coordinate $r/R$ for $LMC\,X-4$. The dot, dash, longdash and solid linestyles represent ${\rho}^{\textit{eff}}+\frac{E^2}{8\pi}$, ${\rho}^{\textit{eff}}+{p}^{\textit{eff}}$, ${\rho}^{\textit{eff}}+3{{p}^{\textit{eff}}}+\frac{E^2}{4\pi}$ and ${\rho}^{\textit{eff}}-{{p}^{\textit{eff}}}+\frac{E^2}{4\pi}$, respectively, whereas the blue, orange, red, green and purple colors represent the cases with $\chi= -0.8$, $\chi= -0.4$, $\chi=0$, $\chi=0.4$ and $\chi=0.8$, respectively. } \label{Fig4}
\end{figure}
%%%%%%%%%%%%%%%%%%%%%%%%%%%%%%%%%%%%%%%%%%%%%%%%%%%%%%%%%%%%%% 

To satisfy all these energy conditions, our system should obey all the inequalities~\eqref{6.1.1}-\eqref{6.1.4} simultaneously. It is evident from Figure~\ref{Fig4} that the configuration is consistent with all the energy conditions.

\subsection{Mass-radius relation}\label{subsec6.2} 

\begin{figure}
\centering
    \includegraphics[width=8cm]{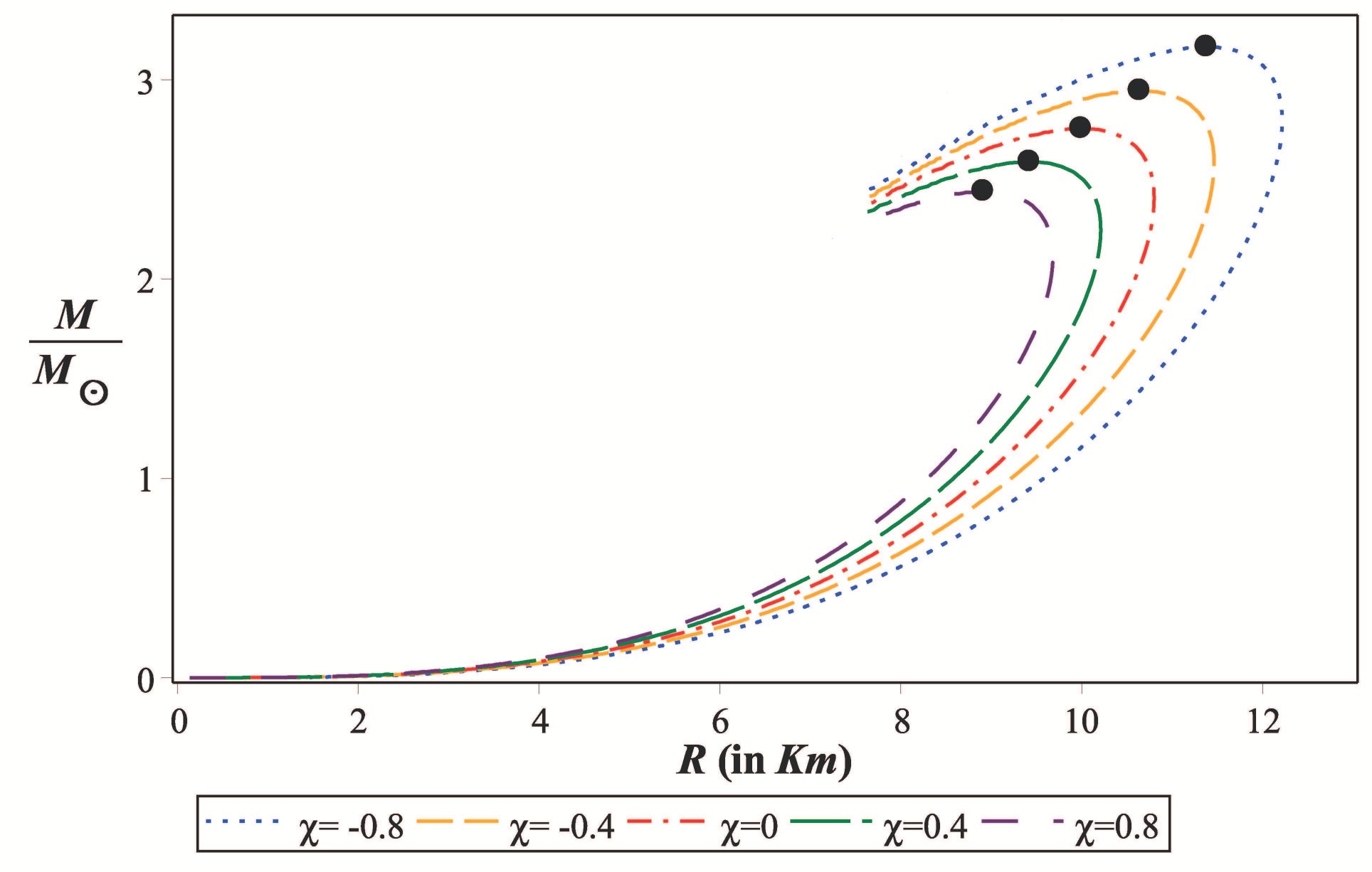}
    \caption{The total mass normalized in the solar mass $(M/M_{\odot})$ as the function of the total radius $(R)$. The solid circles represent the maximal mass of the strange stars.} \label{Fig5}
\end{figure}

Using Eqs.~\eqref{2.3}, \eqref{3.23} and \eqref{3.24} the mass function for our charged stellar system in $f\left(R,\mathcal{T}\right)$ gravity reads
\begin{eqnarray}
m\left(r\right)=\tilde{m}+\frac{\chi}{9\pi}\left[3\,m_0+2\pi B r^3\right],
\end{eqnarray}
where $\tilde{m}=4\pi\int\!\rho {r}^{2}\,{\textrm d}r+\int \!{\frac {q q^\prime }{r}}\,{\textrm d}r$ represents the mass function of the charged SQM distribution and $m_0=4\pi\int\!\rho {r}^{2}\,{\textrm d}r$ denotes the mass function of the uncharged SQM distribution. In the expression of $m(r)$ the term $\frac{\chi}{9\pi}\left[3\,m_0+2\pi B r^3\right]$ represents the mass of the matter distribution and originates due to the coupling between the matter and geometry. We give the total mass normalized with respect to the Solar mass, i.e., $M/M_{\odot}$, with the radius $R$ in Figure~\ref{Fig5} for the chosen specific values of $B$ and $\alpha$ as 83~MeV{fm}$^{-3}$ and 0.0010~{km}$^{-2}$, respectively. Figure~\ref{Fig5} features that, as the values of $\chi$ decrease, the maximal mass and the corresponding radius increase accordingly. 

%%%%%%%%%%%%%%%%%%%%%%%%%%%%%%%%%%%%%%%%%%%%%%%%%%%%%%%%%%%%%%
%%%%%%%%%%%%%%%%%%%%%%%%%%%%%%%%%%%%%%%%%%%%%%%%%%%%%%%%%%%%%%   

%%%%%%%%%%%%%%%%%%%%%%%%%%%%%%%%%%%%%%%%%%%%%%%%%%%%%%%%%%%%%%
\begin{figure}
\centering
    \subfloat{\includegraphics[width=6cm]{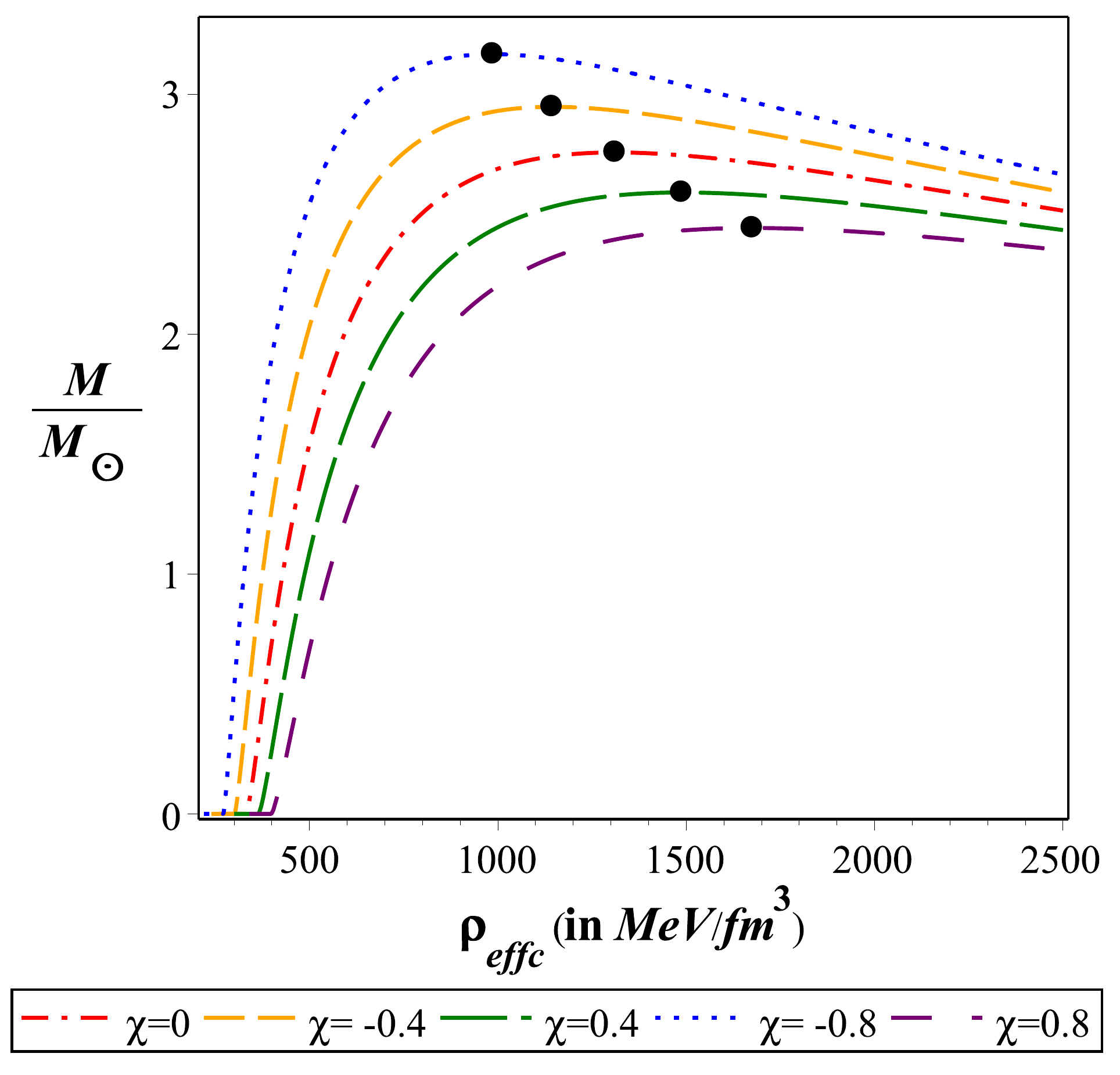}}
    \
    \subfloat{\includegraphics[width=6cm]{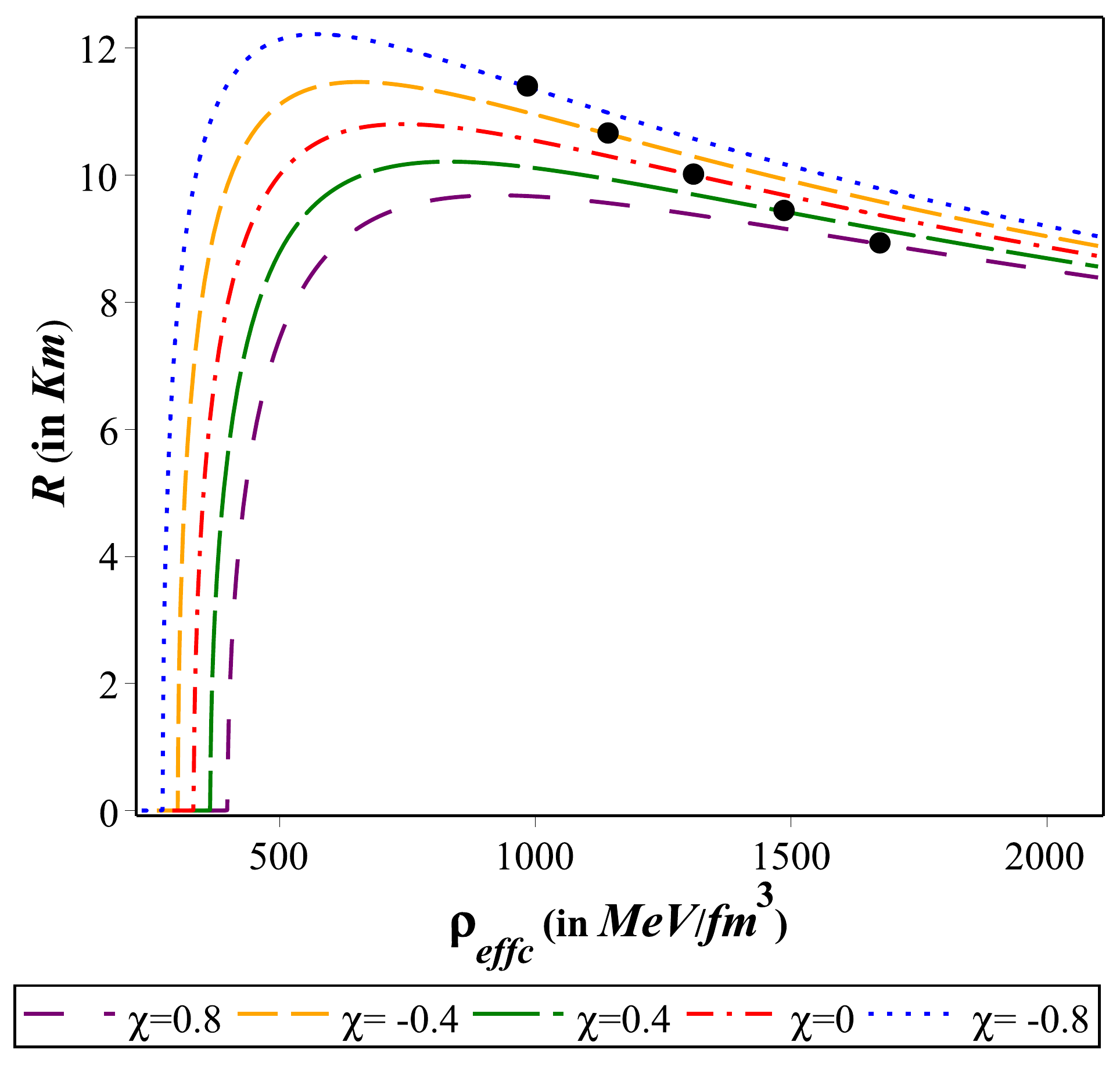}}
    \caption{The (i) $M/M_{\odot}$ and (ii) $R$ as the functions of the central density $\rho_{\textit{effc}}$. The solid circles represent the maximal mass of the strange stars.} \label{Fig6}
\end{figure}
%%%%%%%%%%%%%%%%%%%%%%%%%%%%%%%%%%%%%%%%%%%%%%%%%%%%%%%%%%%%%%  

On the left and right panels of Figure~\ref{Fig6} we show the variation of the total mass $M$ normalized in $M_{\odot}$ and $R$ with respect to the central density $\rho_{\textit{effc}}$, respectively. Figure~\ref{Fig6} reveals that, as the values of $\chi$ decrease, the obtained values of $\rho_{\textit{effc}}$ corresponding to maximal mass $M_{max}$ also decrease gradually. For $\chi=0.8$ and $\chi=-0.8$ we find that $M_{max}=2.442~M_{\odot}$ and $M_{max}=3.166~M_{\odot}$, respectively, whereas the corresponding radii are $R_{Mmax}=8.912$~km and $R_{Mmax}=11.381$~km, respectively. Note that for $\chi=0.8$ and $\chi=-0.8$ the maximal mass points are achieved for $\rho_{\textit{effc}}=2.988\times {10}^{15}$~gm {cc}$^{-1}$ and $\rho_{\textit{effc}}=1.759\times {10}^{15}$~gm {cc}$^{-1}$, respectively. The Figures~\ref{Fig5} and ~\ref{Fig6} suggest that with the decreasing values of $\chi$ the compact stellar system becomes more massive and bigger in size, thus turning the system into a less dense compact stellar object.

\subsection{Stability of the stellar system}\label{subsec6.3}

To examine stability of the ultra dense stellar system, we study (i) modified TOV equation, (ii) causality condition, and (iii) the adiabatic index.

\subsubsection{Modified TOV equation in $f\left(R,\mathcal{T}\right)$ gravity theory}\label{subsubsec6.3.1}
The conservation equation of the energy-momentum tensor in Eqs.~\eqref{3.12} and \eqref{3.25} 
implies the modified form of the TOV equation in $f\left(R,\mathcal{T}\right)$ gravity theory for a charged stellar system as follows:
\begin{eqnarray}\label{6.3.1.1}
&\hspace{-2cm} -\frac{\textrm{d}p}{\textrm{d}r}-\frac{1}{2}\nu_r \left(\rho+p\right)+\frac{\chi}{8\pi+2\chi}\left(\rho^{\prime}-p^{\prime}\right)  +\frac{8\pi}{8\pi+2\chi} \frac{q}{4\pi r^4} \frac{\textrm{d}q}{\textrm{d}r}=0,
\end{eqnarray}
where the terms on the left of Eq.~\eqref{6.3.1.1} represent the hydrodynamic force ($F_h$), the gravitational force ($F_g$), the new kind of force originating due to the coupling between the matter and geometry, which we call the coupling force ($F_c$), and the electric force ($F_e$), respectively. Eq.~\eqref{6.3.1.1} features that $F_h+F_g+F_c+F_e=0$ and shows stability of our system in terms of the equilibrium of forces.

%%%%%%%%%%%%%%%%%%%%%%%%%%%%%%%%%%%%%%%%%%%%%%%%%%%%%%%%%%%%%%
\begin{figure}
\centering
    \includegraphics[width=6cm]{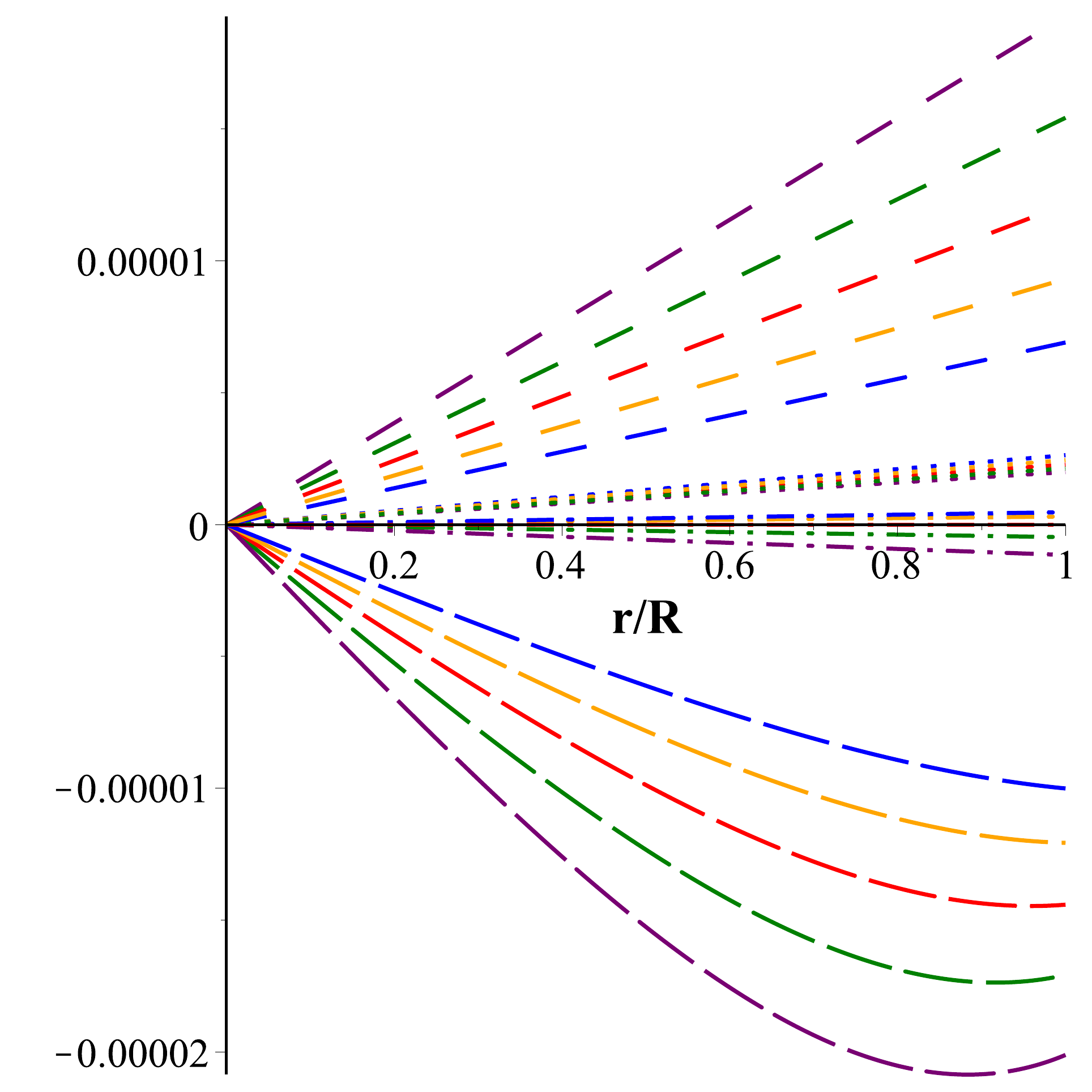}
    \caption{The forces as the functions of the radial coordinate $r/R$ for the strange star candidate $LMC~X-4$. The spacedash, longdash, dashdot and dot linestyle represent $F_h$, $F_g$, $F_c$ and $F_e$, respectively, whereas the blue, orange, red, green and purple colors represent the cases with $\chi=-0.8$, $\chi=-0.4$, $\chi=0$, $\chi=0.4$ and $\chi=0.8$, respectively. } \label{Fig7}
\end{figure}
%%%%%%%%%%%%%%%%%%%%%%%%%%%%%%%%%%%%%%%%%%%%%%%%%%%%%%%%%%%%%%  

In Figure~\ref{Fig7} we show the variation of the forces with the fractional radial coordinate $r/R$. Figure~\ref{Fig7} reveals that for $\chi>0$  the combined effects of $F_g$ and $F_c$ are counterbalanced by the resultant effects of $F_h$ and $F_e$, whereas for $\chi<0$ the effect of $F_g$ is counterbalanced by the combined effects of $F_h$, $F_e$ and $F_c$. Hence, for $\chi>0$ the $F_c$ shows attractive nature and acts along the inward direction, whereas for $\chi<0$ the nature of $F_c$ is repulsive and it acts along the outward direction. 

\subsubsection{Principle of causality}~\label{6.3.2}

It is essential for a physically stable stellar system that the sound speed $v_s$ should be less than the speed of light. The required condition is given by $0<v^2_s<1$ and is known as the principle of causality. For our system, the square of the sound speed is given by
\begin{eqnarray}\label{6.3.2.1}
{v_{{s}}}^{2}=\frac{\textrm{d}p^{\textit{eff}}}{\textrm{d}\rho^{\textit{eff}}}={\frac {\pi }{3 \pi +\chi}}.
\end{eqnarray}

%%%%%%%%%%%%%%%%%%%%%%%%%%%%%%%%%%%%%%%%%%%%%%%%%%%%%%%%%%%%%%
\begin{figure}
\centering
    \includegraphics[width=6cm]{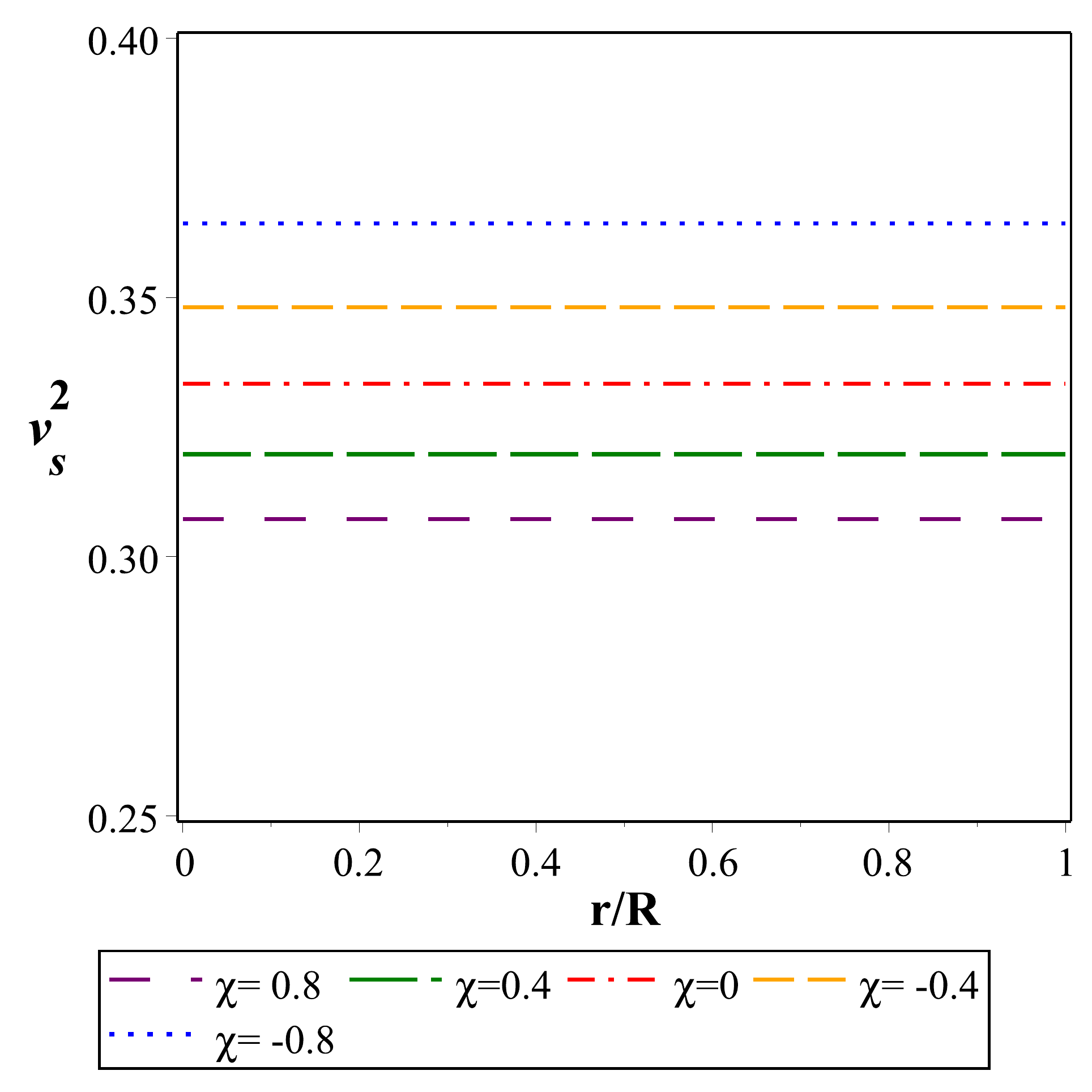}
    \caption{The square of the sound velocity $v^2_s$ as the function of the fractional radial coordinate $r/R$. } \label{Fig8}
\end{figure}
%%%%%%%%%%%%%%%%%%%%%%%%%%%%%%%%%%%%%%%%%%%%%%%%%%%%%%%%%%%%%%  

We give the behaviour of $v^2_s$ in Figure~\ref{Fig8}. It clearly shows that for all  chosen values of $\chi$ our system is well within the inequality $0<v^2_s<1$, which confirms stability of the system under the causality principle.

%%%%%%%%%%%%%%%%%%%%%%%%%%%%%%%%%%%%%%%%%%%%%%%%%%%%%%%%%%%%%%
\begin{figure}
\centering
    \includegraphics[width=6cm]{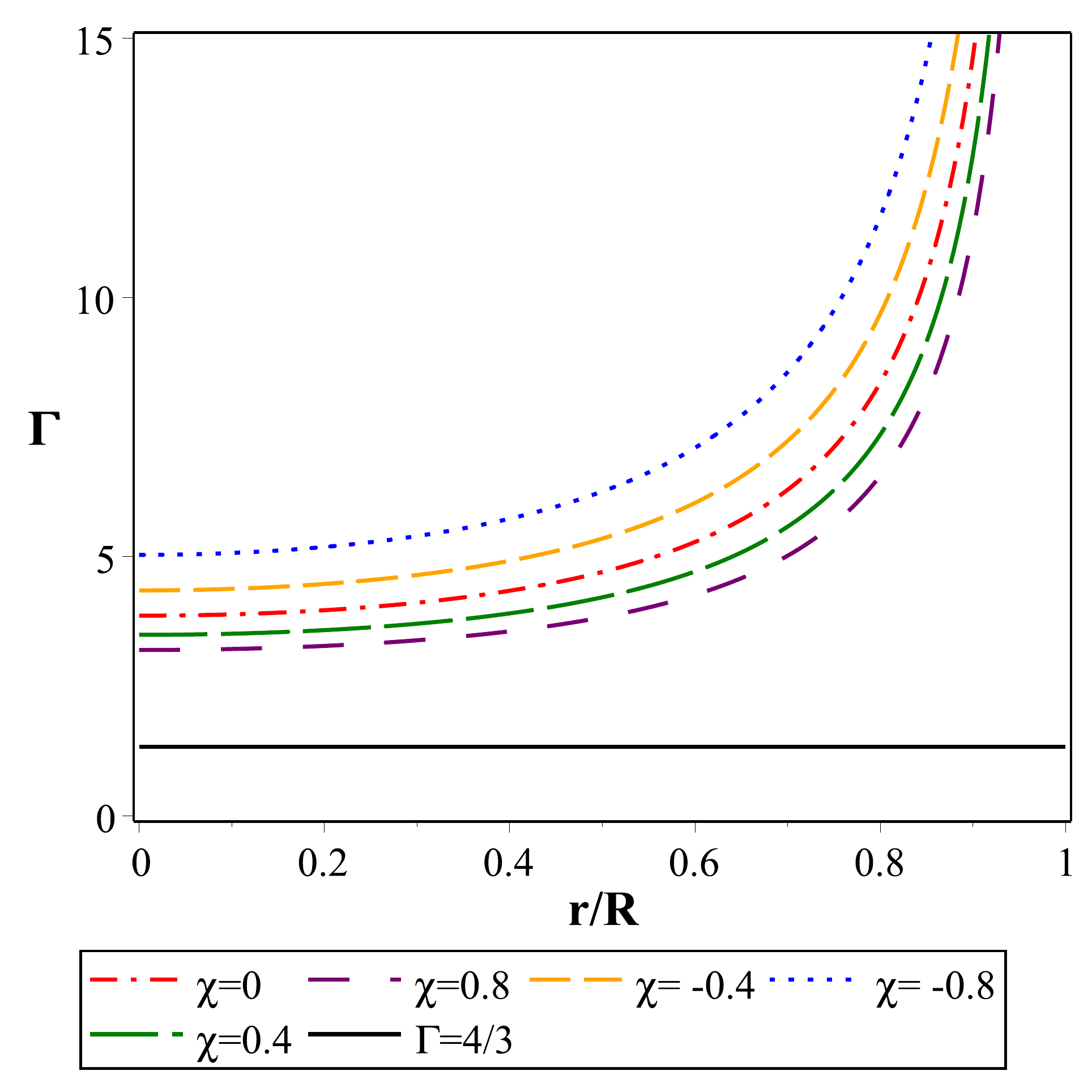}
    \caption{The adiabatic index as the function of the fractional radial coordinate $r/R$. } \label{Fig9}
\end{figure}
%%%%%%%%%%%%%%%%%%%%%%%%%%%%%%%%%%%%%%%%%%%%%%%%%%%%%%%%%%%%%% 

\subsubsection{Adiabatic index}

The study of the adiabatic index is important for the spherically symmetric stellar systems because of the fact that at the given density it characterizes the stiffness of the EOS~\cite{Harrison1965,Haensel2007}. Making a connection between the EOS of the interior matter distribution and the relativistic spherically symmetric static stellar systems, the adiabatic index actually includes all the essential basic characteristics of EOS against instabilities. Chandrasekhar~\cite{Chandrasekhar1964a,Chandrasekhar1964b} 
provided the elegant techniques to examine dynamical stability of the isotropic relativistic stars against an infinitesimal radial adiabatic perturbation. In another work by Heintzmann and Hillebrandt~\citep{Heintzmann1975} it was predicted that in all the interior points the adiabatic index should exceed $4/3$ to achieve a dynamically stellar system. 

For our system the adiabatic index is defined by
\begin{eqnarray}
& \Gamma=\frac{\rho^{\textit{eff}}+p^{\textit{eff}}}{p^{\textit{eff}}} \frac{\textrm{d}p^{\textit{eff}}}{\textrm{d}\rho^{\textit{eff}}}=\frac{1}{{8\Gamma _{{1}}}} \Big[8 {R}^{5}{\pi }^{2}\Gamma _{{1}}-9 \pi {R}^{7}{\alpha
}^{2}- ( 48 B{\pi }^{2}  - ( 9 {\alpha}^{2}{r}^{2}-36 B\chi ) \pi +6 B{\chi}^{2} ) {R}^{5} \nonumber\\
& \hspace{4cm} +80 {r}^{2}\nu_{{1}}{R}^{3} +15 M\pi {R}^{2} -15 M \pi {r}^{2}\Big],
\end{eqnarray}
where $\Gamma _{{1}}=-10 \left( {\frac {9 \pi {R}^{5}{\alpha}^{2}}{80}}+\nu_{{1}}{R}^{3}-\frac{3}{16} M \pi  \right)  \left( R^2-r^2 \right)$.

We present the variation of the parameter $\Gamma$ with respect to the coordinate $r/R$ in Figure~\ref{Fig9}. It indicates that in the present study the value of $\Gamma$ is greater than $4/3$ for all the interior points. This confirms the dynamic stability of the system against infinitesimal radial adiabatic perturbations.

\subsection{Redshift}\label{subsec6.4}

\begin{figure}
\centering
    \includegraphics[width=6cm]{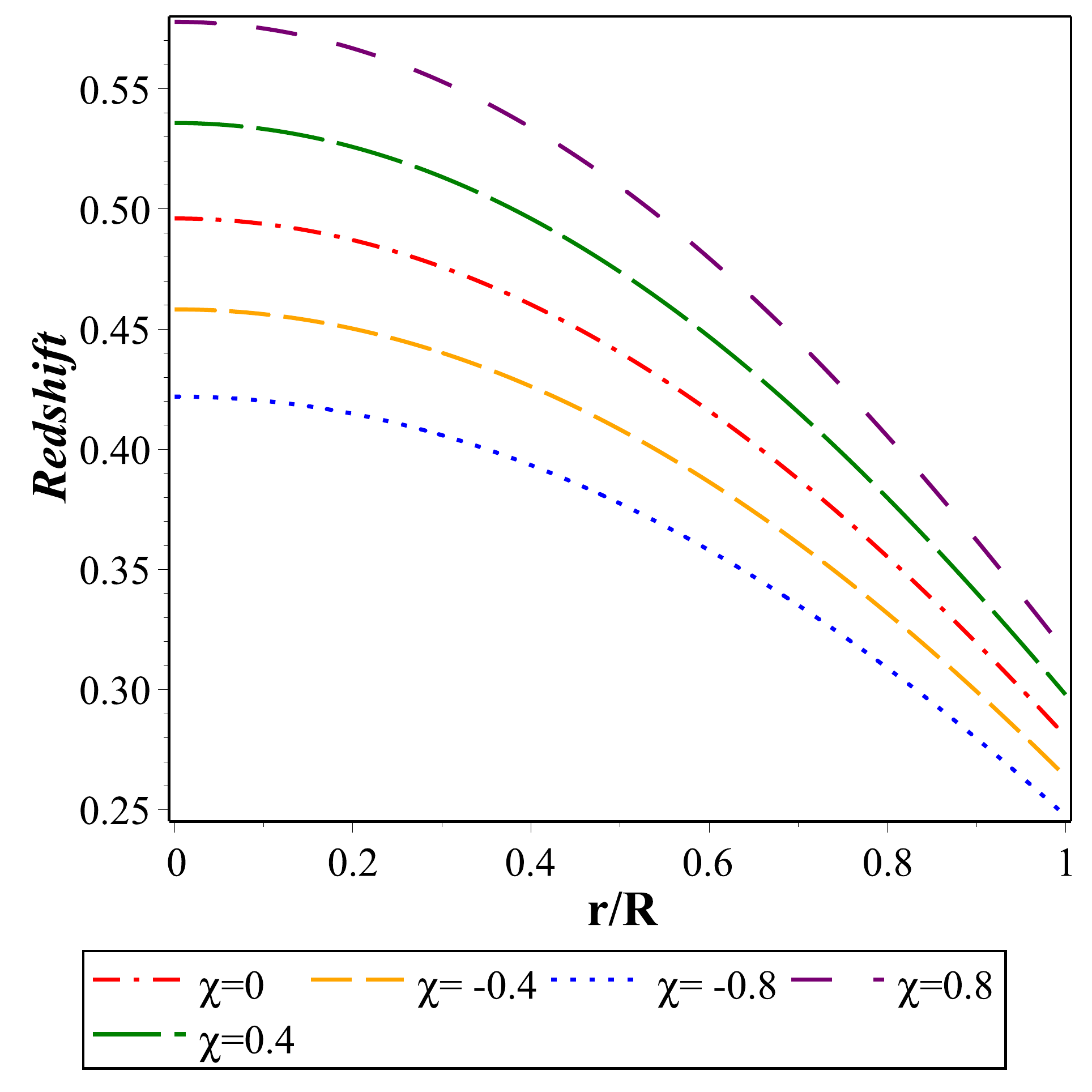}
    \caption{The redshift as the function of the fractional radial coordinate $r/R$. } \label{Fig10}
\end{figure}

Using the basic definition of the gravitational redshift, i.e., $Z_s=\nabla\lambda/\lambda_e=\frac{\lambda_o-\lambda_e}{\lambda_e}$, where $\lambda_o$ is the observed wavelength at a distance $r$,  and $\lambda_e$ is the emitted wavelength from the surface of a spherically symmetric static compact object, we define the surface redshift $Z_s(R)$ for our system as follows:
\begin{eqnarray}
Z_s(R)=e^{-{\frac{\nu(R)}{2}}}-1=\frac{1}{1-2u(R)}-1,
\end{eqnarray}
where $u(r)=m(r)/r$ is the compactification factor of the stellar system. We show the behaviour of the gravitation redshift for the system in Figure~\ref{Fig10}.

%%%%%%%%%%%%%%%%%%%%%%%%%%%%%%%%%%%%%%%%%%%%%%%%%%%%%%%%%%%%%%
%%%%%%%%%%%%%%%%%%%%%%%%%%%%%%%%%%%%%%%%%%%%%%%%%%%%%%%%%%%%%% 

%%%%%%%%%%%%%%%%%%%%%%%%%%%%%%%%%%%%%%%%%%%%%%%%%%%%%%%%%%%%%%%%%%%%%%%%%%%%%%%%%%%%%%%%%%%%%%%%

\begin{table*}[t]
  \centering
    \caption{The physical parameters of the observed candidates of strange stars for $\chi=-0.4$, $\alpha=0.0010 {Km}^{-2} $ and $B=83 MeV/fm^3$.}\label{Table 1}
        \scalebox{0.7}{
\begin{tabular}{ ccccccccccccccccccccccccccc}
\hhline{=========}
Strange  & Observed & Predicted  & $\rho_{\textit{effc}}$  & $p_{\textit{effc}}$ & $Q$ & $E$  & $2M/R$ & $Z_s$  \\ 
Stars &  Mass $(M_{\odot})$ & Radius (Km) & $(gm/{cm}^3)$ & $(dyne/{cm}^2)$ & $(Coloumb)$ & $(V/cm)$ &  &  \\
\hline 
$PSR~J1614-2230$ & $1.97 \pm 0.04$~\cite{Demorest2010} & ${11.041}^{+0.046}_{-0.049}$ & $8.721\times {{10}^{14}}$ & $1.050\times {{10}^{35}}$ & $1.562\times {{10}^{20}}$ &  $1.152\times {{10}^{22}}$  & $0.526$ & $0.452$ \\ 
\ 
$Vela~X-1$ & $1.77 \pm 0.08$~\cite{Rawls2011} & ${10.777}^{+0.112}_{-0.121}$ & $8.161\times {{10}^{14}}$ & $8.746\times {{10}^{34}}$ & $1.452\times {{10}^{20}}$  &  $1.125\times {{10}^{22}}$   & $0.484$ & $0.392$ \\ 
\ 
$4U~1608-52$ & $1.74\pm 0.14$~\cite{guver2010a} & ${10.732}^{+0.196}_{-0.222}$ & $8.089\times {{10}^{14}}$ & $8.522\times {{10}^{34}}$ & $1.434\times {{10}^{20}}$    &  $1.120\times {{10}^{22}}$   &     $0.478$ & $0.384$ \\ 
\ 
$PSR~J1903 + 327$ & $1.667 \pm 0.021$~\cite{Freire2011} & ${10.620}^{+0.033}_{-0.034}$ & $7.912\times {{10}^{14}}$ & $7.967\times {{10}^{34}}$ & $1.390\times {{10}^{20}}$   &  $1.108\times {{10}^{22}}$   &    $0.463$ & $0.365$ \\ 
\
$4U~1820-30$ & $1.58\pm 0.06$~\cite{guver2010b} & ${10.476}^{+0.100}_{-0.105}$ & $7.719\times {{10}^{14}}$ & $7.364\times {{10}^{34}}$ & $1.334\times {{10}^{20}}$ &  $1.093\times {{10}^{22}}$   &     $0.445$ & $0.342$ \\ 
\
$Cen~X-3$ & $1.49 \pm 0.08$~\cite{Rawls2011}  & ${10.317}^{+0.142}_{-0.152}$ & $7.528\times {{10}^{14}}$ & $6.767\times {{10}^{34}}$ &  $1.274\times {{10}^{20}}$     &  $1.077\times {{10}^{22}}$   &     $0.426$ & $0.320$ \\ 
\
$EXO~1785-248$ & $1.3\pm 0.2$~\cite{ozel2009} & ${9.940}^{+0.395}_{-0.462}$ & $7.172\times {{10}^{14}}$ & $5.653\times {{10}^{34}}$ & $1.139\times {{10}^{20}}$     &  $1.037\times {{10}^{22}}$   &     $0.386$ & $0.276$ \\ 
\ 
$LMC~X - 4$ & $1.29 \pm 0.05$~\cite{Rawls2011} & ${9.919}^{+0.105}_{-0.110}$ & $7.152\times {{10}^{14}}$ & $5.592\times {{10}^{34}}$ &  $1.132\times {{10}^{20}}$   &  $1.035\times {{10}^{22}}$   &     $0.384$ & $0.274$ \\ 
\ 
$SMC~X - 1$ & $1.04\pm 0.09$~\cite{Rawls2011} & ${9.324}^{+0.228}_{-0.246}$ & $6.747\times {{10}^{14}}$ & $4.323\times {{10}^{34}}$ &  $9.405\times {{10}^{19}}$    &  $9.729\times {{10}^{21}}$   &     $0.329$ & $0.221$ \\ 
\ 
$SAX~J1808.4-3658$ & $0.9\pm 0.3$~\cite{Elebert2009} & ${8.932}^{+0.786}_{-1.044}$ & $6.544\times {{10}^{14}}$ & $3.688\times {{10}^{34}}$ &  $8.268\times {{10}^{19}}$   &  $9.320\times {{10}^{21}}$   &     $0.297$ & $0.193$ \\ 
\ 
$4U~1538-52$ & $0.87\pm 0.07$~\cite{Rawls2011} & ${8.842}^{+0.208}_{-0.222}$ & $6.499\times {{10}^{14}}$ & $3.547\times {{10}^{34}}$ &  $8.021\times {{10}^{19}}$    &  $9.226\times {{10}^{21}}$   &     $0.290$ & $0.187$ \\ 
\ 
$HER~X-1$ & $0.85\pm 0.15$~\cite{Abubekerov2008} & ${8.780}^{+0.437}_{-0.505}$ & $6.472\times {{10}^{14}}$ & $3.464\times {{10}^{34}}$ &  $7.853\times {{10}^{19}}$   &  $9.162\times {{10}^{21}}$   &     $0.286$ & $0.183$ \\ 
\hhline{=========} 
\end{tabular}  }
  \end{table*}

%%%%%%%%%%%%%%%%%%%%%%%%%%%%%%%%%%%%%%%%%%%%%%%%%%%%%%%%%%%%%%%%%%%%%%%%%%%%%%%%%%%%%%%%%%%%%%%%
  
%%%%%%%%%%%%%%%%%%%%%%%%%%%%%%%%%%%%%%%%%%%%%%%%%%%%%%%%%%%%%%%%%%%%%%%%%%%%%%%%%%%%%%%%%%%%%%%%

\begin{table*}
  \centering
    \caption{The physical parameters of the strange star candidate $LMC~X-4$ at the different values of $\chi$, $\alpha=0.0010 {Km}^{-2} $ and $B=83 MeV/fm^3$.}\label{Table 2}
        \scalebox{0.75}{
\begin{tabular}{ ccccccccccccccccccccccccccc}
\hhline{=========} 
 Values of $\chi$ & Predicted  & $\rho_{\textit{effc}}$ & $\rho_{\textit{eff0}}$ & $p_{\textit{effc}}$ & $Q$ & $E$ & $2M/R$ & $Z_s$ \\
  &  radius (Km) & $(gm/{cm}^3)$ & $(gm/{cm}^3)$ & $(dyne/{cm}^2)$ & $(Coloumb)$ & $(V/cm)$ &  & \\ 
 \hline 
 $-0.8$ & ${10.319}^{+0.113}_{-0.117}$ & $6.154\times {{10}^{14}}$ & $4.836\times {{10}^{14}}$ & $4.314\times {{10}^{34}}$ & $1.275\times {{10}^{20}}$ & $1.077\times {{10}^{22}}$ & $0.369$ & $0.259$ \\ 
 \ 
 $-0.4$ & ${9.919}^{+0.105}_{-0.110}$ & $7.153\times {{10}^{14}}$ & $5.365\times {{10}^{14}}$ & $5.593\times {{10}^{34}}$ & $1.132\times {{10}^{20}}$ & $1.035\times {{10}^{22}}$ & $0.384$ & $0.274$ \\ 
 \ 
 $0$ & ${9.548}^{+0.098}_{-0.103}$ & $8.259\times {{10}^{14}}$ & $5.918\times {{10}^{14}}$ & $7.013\times {{10}^{34}}$ & $1.010\times {{10}^{20}}$ & $9.963\times {{10}^{21}}$ & $0.399$ & $0.290$ \\ 
 \ 
 $0.4$ & ${9.201}^{+0.091}_{-0.096}$ & $9.486\times {{10}^{14}}$ & $6.496\times {{10}^{14}}$ & $8.593\times {{10}^{34}}$ & $9.038\times {{10}^{19}}$ & $9.601\times {{10}^{21}}$ & $0.414$ & $0.306$ \\ 
 \ 
 $0.8$ & ${8.877}^{+0.084}_{-0.089}$ & $1.085\times {{10}^{15}}$ & $7.097\times {{10}^{14}}$ & $1.036\times {{10}^{35}}$ & $8.116\times {{10}^{19}}$ & $9.263\times {{10}^{21}}$ & $0.429$ & $0.323$ \\ 
 \hhline{=========} 
\end{tabular}  }
  \end{table*}

%%%%%%%%%%%%%%%%%%%%%%%%%%%%%%%%%%%%%%%%%%%%%%%%%%%%%%%%%%%%%%%%%%%%%%%%%%%%%%%%%%%%%%%%%%%%%%%% 

%%%%%%%%%%%%%%%%%%%%%%%%%%%%%%%%%%%%%%%%%%%%%%%%%%%%%%%%%%%%%%%%%%%%%%%%%%%%%%%%%%%%%%%%%%%%%%%%

\begin{table*}
  \centering
    \caption{The physical parameters of the strange star candidate $LMC~X-4$ at the different values of $\alpha$, $B=83~MeV/{fm}^3 $ and $\chi=-0.4$.}\label{Table 3}
        \scalebox{0.75}{
\begin{tabular}{ ccccccccccccccccccccccccccc}
\hhline{=========} 
  Values of $\alpha$ & Predicted  & $\rho_{\textit{effc}}$  & $p_{\textit{effc}}$ & $Q$ & $E$  & $2M/R$ & $Z_s$   \\ 
                     & Radius (Km) & $(gm/{cm}^3)$ & $(dyne/{cm}^2)$ & $(Coloumb)$ & $(V/cm)$ &   &   \\
  \hline
  $0$ & ${9.888}^{+0.103}_{-0.107}$ & $7.776\times {{10}^{14}}$ & $7.542\times {{10}^{34}}$ & $1.122\times {{10}^{20}}$ & $1.032\times {{10}^{22}}$ & $0.385$ & $0.275$ \\ 
  \
  $0.0005$ & ${9.896}^{+0.103}_{-0.108}$ & $7.621\times {{10}^{14}}$ & $7.056\times {{10}^{34}}$ & $1.124\times {{10}^{20}}$ & $1.033\times {{10}^{22}}$ & $0.385$ & $0.275$ \\ 
  \
  $0.0010$ & ${9.919}^{+0.105}_{-0.110}$ & $7.153\times {{10}^{14}}$ & $5.593\times {{10}^{34}}$ & $1.132\times {{10}^{20}}$ & $1.035\times {{10}^{22}}$ & $0.384$ & $0.274$ \\ 
  \
  $0.0015$ & ${9.958}^{+0.109}_{-0.113}$ & $6.369\times {{10}^{14}}$ & $3.142\times {{10}^{34}}$ & $1.146\times {{10}^{20}}$ & $1.039\times {{10}^{22}}$ & $0.382$ & $0.272$ \\ 
  \hhline{=========} 
\end{tabular}  }
  \end{table*}

%%%%%%%%%%%%%%%%%%%%%%%%%%%%%%%%%%%%%%%%%%%%%%%%%%%%%%%%%%%%%%%%%%%%%%%%%%%%%%%%%%%%%%%%%%%%%%%%

%%%%%%%%%%%%%%%%%%%%%%%%%%%%%%%%%%%%%%%%%%%%%%%%%%%%%%%%%%%%%%%%%%%%%%%%%%%%%%%%%%%%%%%%%%%%%%%%

\begin{table*}
  \centering
    \caption{The physical parameters of the strange star candidate $LMC~X-4$ at the the different values of $B$, $\alpha=0.0010 {Km}^{-2} $ and $\chi=-0.4$.}\label{Table 4}
        \scalebox{0.75}{
\begin{tabular}{ ccccccccccccccccccccccccccc}
\hhline{=========} 
 Values of $B$ & Predicted  & $\rho_{\textit{effc}}$  & $\rho_{\textit{eff0}}$  & $p_{\textit{effc}}$ & $Q$ & $E$  & $2M/R$ & $Z_s$ \\
  &  Radius (Km)  &  $(gm/{cm}^3)$ &  $(gm/{cm}^3)$ &  $(dyne/{cm}^2)$ & $(Coloumb)$ & $(V/cm)$ &  &   \\ 
 \hline 
 $60$ & ${11.162}^{+0.125}_{-0.130}$ & $4.583\times{{10}^{14}}$ & $3.878\times{{10}^{14}}$ & $2.203\times{{10}^{34}}$ &   $1.614\times {{10}^{20}}$ & $1.165\times {{10}^{22}}$  & $0.341$ & $0.232$ \\ 
 \ 
 $80$ & ${10.054}^{+0.108}_{-0.112}$ & $6.810\times{{10}^{14}}$ & $5.171\times{{10}^{14}}$ & $5.126\times{{10}^{34}}$ &    $1.179\times {{10}^{20}}$ & $1.049\times {{10}^{22}}$  &   $0.378$ & $0.268$ \\ 
 \
 $90$ & ${9.628}^{+0.101}_{-0.105}$ & $7.965\times{{10}^{14}}$ & $5.818\times{{10}^{14}}$ & $6.718\times{{10}^{34}}$ &   $1.035\times {{10}^{20}}$ & $1.005\times {{10}^{22}}$  &  $0.395$ & $0.286$ \\ 
 \hhline{=========} 
\end{tabular}  }
  \end{table*}

%%%%%%%%%%%%%%%%%%%%%%%%%%%%%%%%%%%%%%%%%%%%%%%%%%%%%%%%%%%%%%%%%%%%%%%%%%%%%%%%%%%%%%%%%%%%%%%% 

\section{Discussion and conclusion}\label{sec7}

Based on the simplified form of $f\left(R,\mathcal{T}\right)$ function~\cite{Harko2011}, for the first time in the literature to our best knowledge, we present a model for the charged ultra dense strange quark stars, where the modified $f\left(R,\mathcal{T}\right)$ gravity leads to significant effects on the stellar structure. Our system is governed by the simplified phenomenological MIT bag model EOS, viz., $p=\left(\rho-4\,B\right)/3$, and the electrical charge distribution $q\left(r\right)=Q\left(r/R\right)^3$~\cite{Felice1995}. In Eq.~\eqref{3.5} we give the Einstein-Maxwell field equation for $f\left(R,\mathcal{T}\right)$ gravity theory, and obtain the covariant divergence of energy-momentum tensor in Eq.~\eqref{3.8}. Assuming  $f\left(R,\mathcal{T}\right)=R+2\chi\mathcal{T}$~\cite{Harko2011} we find the new kind of fluid originating from the interaction between the matter and geometry, which is evident from Eqs.~\eqref{3.9} and \eqref{3.10}. Importantly, it is shown for our system by considering the energy-momentum tensor of the effective matter distribution,  
$T^{eff}_{\mu\nu}$, that the conservation of the energy-momentum tensor is achieved - see Eq.~\eqref{3.11} --- while the equivalence principle is still valid for our system in the framework of the $f\left(R,\mathcal{T}\right)$ gravity theory. We consider $LMC~X-4$ of mass $1.29~M_{\odot}$~\cite{Rawls2011} as the representative of the candidates of strange stars, with the reference values $B=83~MeV/{fm}^3$~\cite{Rahaman2014} and $\alpha=0.0010~km^{-2}$. We give our results for the several different parametric values of $\chi$, viz., $\chi=-0.8, -0.4, 0, 0.4$ and $0.8$.

On the left and right panels of Figure~\ref{Fig1} we give the profiles of the metric potentials $e^{\nu}$ and $e^{\lambda}$ with respect to the fractional radial coordinate $r/R$, respectively. Both metric potentials have finite values at the centre and  monotonically increase from the centre to the surface. The variations of the effective energy density $\rho^{\textit{eff}}$ and pressure $p^{\textit{eff}}$ with $r/R$ are shown on the left and right panels of Figure~\ref{Fig2}, which features both $\rho^{\textit{eff}}$ and $p^{\textit{eff}}$ having their maximal values at the centre and gradually decreasing to reach their minimum value at the surface, which validates the physical validity of the achieved solutions. Interestingly, both Figures~\ref{Fig1} and \ref{Fig2} confirm that our system is free from any type of singularities, either geometrical or physical ones.

Based on the prediction of Ray et al.~\cite{Ray2008}, the profiles of the electric charge distribution function $q\left(r\right)$ and electrical energy density $E^2\left(r\right)/8\pi$ are shown on the left and right panels of Figure~\ref{Fig3}, respectively, which clearly shows that both reach their 
minima and vanish at the centre, while both increase gradually throughout the interior of the stellar system to attain their maximal values at the surface. The total charge $Q$ and the associated electric field $E$ have values of the order of ${10}^{20-19}$ and ${10}^{22-21}$, respectively. We find that, as the values of $\alpha$ increase, the stellar objects under consideration become larger in size and less dense.

%%%%%%%%%%%%%%%%%%%%%%%%%%%%%%%%%%%%%%%%%%%%%%%%%%%%%%%%%%%%%%
\begin{figure}[htp!]
\centering
    \includegraphics[width=6cm]{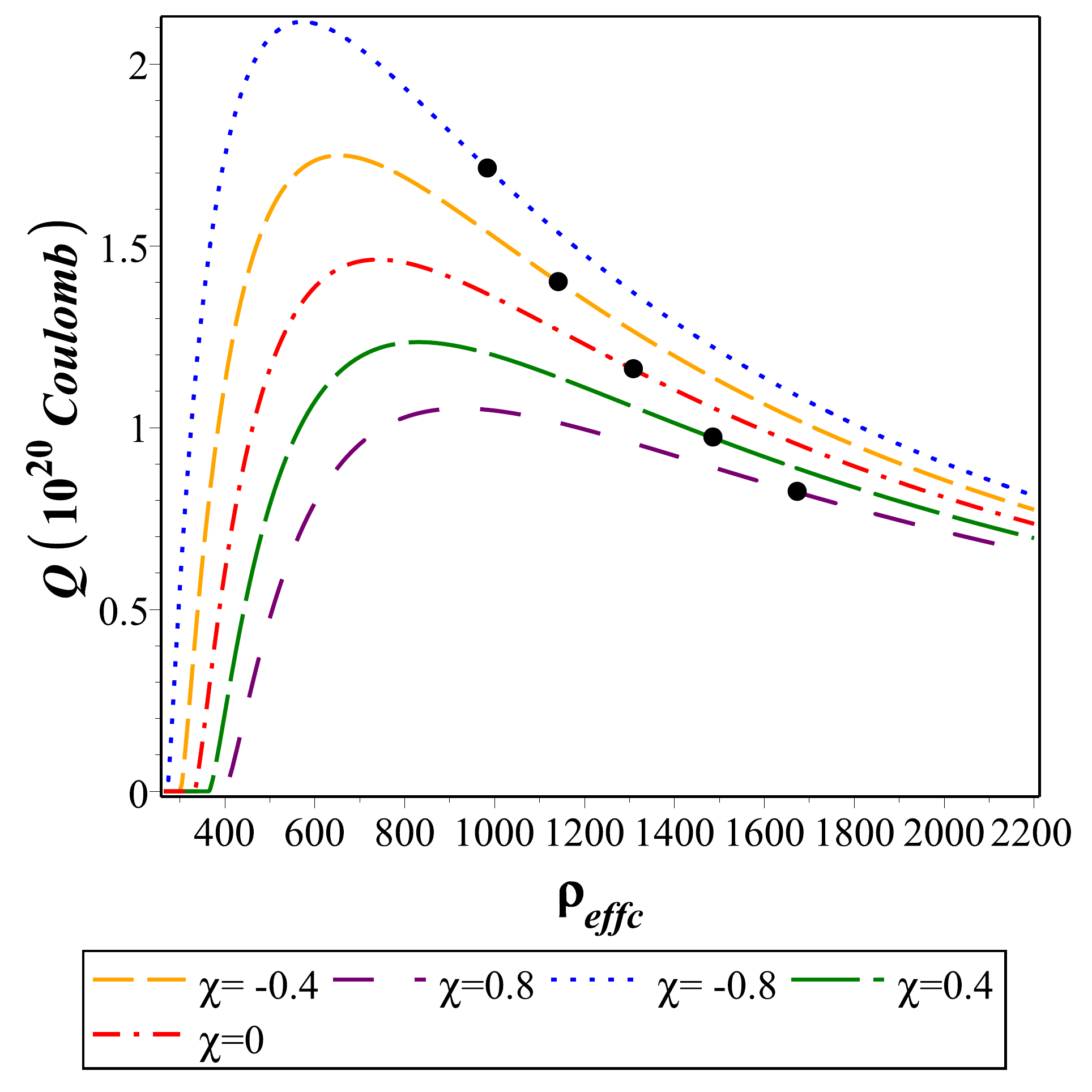}
    \caption{The total charge $Q$ as the function of the central density $\rho_{\textit{effc}}$. The solid circles represent the alues of $\rho_{\textit{effc}}$ corresponding to the maximual mass of the strange stars. } \label{Fig11}
\end{figure}
%%%%%%%%%%%%%%%%%%%%%%%%%%%%%%%%%%%%%%%%%%%%%%%%%%%%%%%%%%%%%% 

Further, we tested the physical viability of the achieved solutions by studying the energy conditions, the mass-radius relation and stability of the stellar systems, etc. Figure~\ref{Fig4} shows that the system satisfies all the inequalities shown in Eqs.~\eqref{6.1.1}-\eqref{6.1.4}, and, hence, is consistent with the energy conditions. We present the modified form of the conservation equation of the system in Eqs.~\eqref{3.12} and~\ref{6.3.1.1}. The profiles of the forces at some different values of $\chi$ are shown in Figure~\ref{Fig7}, which indicates that for $\chi<0$ the coupling force $F_c$ is repulsive in nature and acts along the outward direction to counterbalance the inward gravitational force $F_g$ along with the other repulsive forces, viz., $F_h$ and $F_e$. However, $F_c$ shows attractive nature for $\chi>0$ and acts along the inward direction like $F_g$ to counterbalance the outward combined force by $F_h$ and $F_e$. In both cases, the system is stable in terms of equilibrium of the forces as $F_h+F_e+F_c+F_g=0$. The model is also consistent with the causality condition as $v^2_s$ satisfies the inequality $0<v^2_s<1$ that is manifest in Figure~\ref{Fig8}. In Figure~\ref{Fig9} we show the variation of the adiabatic index $\Gamma$ with the fraction radial coordinate $r/R$, which validates dynamic stability of the system against the infinitesimal radial adiabatic perturbations in all the interior points with  $\Gamma>4/3$. Finally, the  behaviour of the redshift function is shown in Figure~\ref{Fig10}. 

Our study of the mass-radius relation is important for the compact stellar configurations and shows the maximal bound for the stellar objects. In Figure~\ref{Fig5} we present the variation of total mass $M$ (normalized in the Solar mass $M_{\odot}$) with the total radius $R$ for different  parametric values of $\chi$. In the present $f(R,\mathcal{T})$ gravity model we find from Figure~\ref{Fig5} that due to $\chi= -0.8$ the obtained values of the maximal mass and the corresponding radius are $14.860~\%$ and $13.871~\%$ higher than their corresponding values in GR, whereas $29.675~\%$ and $27.704~\%$ higher than the corresponding values in the case of $\chi=0.8$. We show the behaviour of $M$ and $R$ with the effective central density $\rho_{\textit{effc}}$ in Figure~\ref{Fig6}, which implies that the maximal mass bound for $\chi=-0.8$ is achieved for $\rho_{\textit{effc}}=7.648 \rho_{nuclear}$, where $\rho_{nuclear}$ is the normal nuclear density. The obtained value of $\rho_{\textit{effc}}$ corresponding to the maximal mass point for $\chi=-0.8$ is $32.922~\%$ and $69.848~\%$ less than its value in the case of GR and $\chi=0.8$, respectively. Hence, Figures~\ref{Fig5} and \ref{Fig6} clearly indicate that, as the values of $\chi$ increase, the stellar configurations in $f(R,\mathcal{T})$ gravity become massive and larger in size turning it into a less compact stellar object. 

It is known for spherically symmetric static stellar systems that the necessary and sufficient condition for stability is $dM/d\rho_{\textit{effc}}>0$~\cite{Harrison1965}. Figure~\ref{Fig6} (the left panel) confirms that the $f(R,\mathcal{T})$ gravity model in the present study is fully consistent with the stability condition $dM/d\rho_{\textit{effc}}>0$ up to the maximal mass point $M=M_{max}$. In Figure~\ref{Fig11} we present the variation of the total charge $Q$ with $\rho_{\textit{effc}}$ which shows that, as values of $\chi$ decrease, the maximal values of $Q$ increase gradually. This  clearly shows that the coupling between the matter and geometry has the significant effect on the charged stellar configuration. We find that increasing values of $\alpha$ and decreasing values of $\chi$ gradually push the maximal upper limit of the compact stellar objects beyond the standard values of TOV limit.

Considering the observed mass values of the strange stars candidates in Table~\ref{Table 1}, we predict the physical parameters of the stellar objects, viz., $R$, $\rho_{\textit{effc}}$, the surface density ($\rho_{\textit{eff0}}$), the central pressure ($p_{\textit{effc}}$), $Q$, etc., for $B$=83~MeV fm$^{-3}$, $\chi=-0.4$ and $\alpha$=0.0010~Km$^{-2}$. In Table~\ref{Table 2} we show how the same physical parameters get changed with the change of the  parameter $\chi$. We find that, as the values of $\chi$ decrease from $0.8$ to $-0.8$, the physical parameters $\rho_{\textit{effc}}$, $\rho_{\textit{eff0}}$, $p_{\textit{effc}}$, $2M/R$ and $Z_s$ decrease gradually, whereas $R$, $Q$ and $E$ increase accordingly. We observe the same characteristics of the physical parameters with the change of the parameter $\alpha$ in Table~\ref{Table 3}: as the values of $\alpha$ increase from $0$ to $0.0015$, the predicted values of $\rho_{\textit{effc}}$, $p_{\textit{effc}}$, $2M/R$ and $Z_s$ decrease gradually, whereas $R$, $Q$ and $E$ increase in the same way as is shown in Table~{\ref{Table 2}. We also show how the different physical parameters change with the different values of $B$ in table~\ref{Table 4}. Hence, when following the results in Tables~\ref{Table 2} and \ref{Table 3} and in Figures~\ref{Fig3} and \ref{Fig11}, we conclude that $\chi$ is inversely related to $Q$.

The success of GR in the weak gravitational field regime (e.g. the Solar system tests and laboratory experiments, etc.) is out of the question. However, GR faces challenges in explaining the strong gravity field regime (e.g. in the early Universe, in the regions near to ultra-dense compact stars and black holes). The recent observation of the highly massive pulsars with masses $2.27^{+0.17}_{-0.15}~M_{\odot}$~\cite{Linares2018} in the binary $PSR~J2215+5135$ is a serious setback to the standard TOV mass limit of the neutron stars. Besides, the Chandrasekhar mass limit hardly can explain the surprising discovery of the highly over-luminous SNeIa (supernovae, like $SN~2003fg$, $SN~2006gz$, $SN~2007if$ and $SN~2009dc$)~\cite{Howell2006,Scalzo2010} which not only confirms the high Ni-mass but also invokes the possibility of super-Chandrasekhar white dwarfs of masses $2.1-2.8~M_{\odot}$ as the progenitors of the peculiar highly over-luminous SNela~\cite{Howell2006,Hicken2007,Yamanaka2009,Scalzo2010,Silverman2011,Taubenberger2011}. In the work of Harko and collaborators~\cite{Mak2004,Kovacs2009} the strange stars are predicted to have higher maximal masses compared to the neutron stars, so that the values of the maximal masses for the quark stars can reach $6-7~M_{\odot}$. This may imply that some stellar mass black holes are actually the strange stars. The $f(R,\mathcal{T})$ gravity model reveals that the maximal mass limit for strange stars increases with the increasing values of $\alpha$ and decreasing values of $\chi$. Hence, for the appropriate choice of $\chi$ and $\alpha$, our 
$f(R,\mathcal{T})$ gravity model can be considered as the suitable way to explain the above mentioned super-Chandrasekhar white dwarfs, massive pulsars and magnetars, etc. Though the main goal of this investigation is a study of the effects of $f(R,\mathcal{T})$ gravity theory on the charged strange stars, it would be interesting to examine whether our $f(R,\mathcal{T})$ gravity model is also applicable to the above mentioned unexplained astrophysical situations.

In a nutshell, our model of the strange stars survives all the critical physical tests conducted in the present investigation and successfully explains the effects of electric charge in the context of the modified $f(R,\mathcal{T})$ gravity.

\section*{ACKNOWLEDGMENTS}

SR is grateful to the Inter-University Centre for Astronomy and Astrophysics (IUCAA), Pune, India for providing Associateship Programme under which a part of this work was carried out. Another part of this work was completed by DD when he was visiting the IUCAA, and DD gratefully acknowledges warm hospitality there. The work of SVK was supported in part by the Competitiveness Enhancement Program of Tomsk Polytechnic University in Russia, and the World Premier International Research Center Initiative (WPI Initiative), MEXT, Japan. The work on astrophysical probes for BSM models by MK was supported by a grant of Russian Science Foundation (project N-18-12-00213).

\section*{APPENDIX}

The constants used in Eqs.~\eqref{3.28}-\eqref{3.31} are given by
\begin{eqnarray}\label{appen1}
& & \nu_{{1}}= \left( \pi +\frac{\chi}{4} \right)  \left( \pi +\frac{\chi}{2} \right) B, \\ \label{appen2}
& & \nu_{{2}}=\Bigg[\frac {9{R}^{10}{\pi }^{2}{\alpha}^{4}}{256}+\frac{3}{8}\,{\alpha}^{2}\nu_{{1}}\pi {R}^{8}+ \left( \nu^2_{{1}}+\frac{1}{32}{\pi }^{2}{\alpha}^{2} \right) {R}^{6}  -{\frac {15\,M{R}^{5}{\pi }^{2}{\alpha}^{2}}{128}}+\frac{1}{4}\nu_{{1}}\pi \,{R}^{4}\nonumber \\
& & \hspace{2cm} -\frac{5}{8}M\nu_{{1}}\pi \,B{R}^{3}-{\frac {3\,MR{\pi }^{2}}{64}}  +{\frac {25\,{M}^{2}{\pi }^{2}}{256}}\Bigg]^{\frac{1}{2}} R^2,   \\ \label{appen3}
& & \nu_{{3}}=-\frac{1}{8}\nu_{{2}} \Big[ -12\,{\pi }^{2}{R}^{5}{\alpha}^{2}-\pi \,{R}^{5}{\alpha}^{2}\chi-128\,\pi \,{R}^{3}\nu_{{1}} -16\,{R}^{3}\chi\,\nu_{{1}}+24\,M{\pi }^{2}+3\,M\pi \,\chi \Big],~~~~~  \\ \label{appen4}
& & \nu_{{4}}=32{R}^{2}\left( \pi +\frac{\chi}{4} \right)\Bigg[ {\frac {9\,{R
}^{10}{\pi }^{2}{\alpha}^{4}}{512}}+\frac{1}{4}\,\pi \, \left( \pi +\frac{3}{16}\,\chi
 \right)  \big( \pi  +\frac{\chi}{2} \big) B{\alpha}^{2}{R}^{8}+\nu_{{1}}
 \left( \pi +\frac{\chi}{2} \right)  \nonumber \\
& & \hspace{0.8cm} \Big( \pi +\frac{\chi}{8} \Big) B{R}^{6} -{\frac {3\,M{R}^{5}{\pi }^{2}{\alpha}^{2}}{64}} -\frac {11}{32} \pi B \,M \left( \pi +\frac{\chi}{2} \right) \left( \pi +\frac{2}{11}\chi \right) {R}^{3}+{\frac {15\,{M}^{2}{\pi }^{2}}{512}} \Bigg],  \\ \label{appen5}
& & \nu_{{5}}= \frac {1}{24\nu_{{2}} \left( \pi +\frac{\chi}{3} \right) 
 \left( \frac{1}{8}\pi{R}^{5}{\alpha}^{2}+\nu_{{1}}{R}^{3}-\frac{3}{16}M\pi\right) }.
\end{eqnarray}

\vspace{1.0cm}

\end{document}